   \documentstyle[editedvolume]{crckapb} 

    \begin{opening} 
    \title{Ultraviolet sky surveys} 
    \subtitle{Instruments, findings, and prospects} 
 
    \author{Noah Brosch}

    \institute{Space Telescope Science Institute \\ 3700 San Martin Drive \\
Baltimore MD 21218, U.S.A.}

    \end{opening} 
 
    \runningtitle{UV sky surveys} 

    \def\etal{{\it et al.\ }}

\begin{document} 


     \maketitle 

  \footnote {On sabbatical leave from the 
Department of Astronomy and Astrophysics and the Wise 
    Observatory, 
     School of Physics and Astronomy, 
    Raymond and Beverly Sackler Faculty of Exact Sciences, 
     Tel Aviv University, Tel Aviv 69978, Israel}

        \begin{abstract} 
 
    I review the development of UV and EUV astronomy, covering the spectral 
    range from 5 to 300 nm, with emphasis on sky surveys for discrete 
    sources. I discuss studies which resulted in lists of sources observed 
    by imaging and deliberately omit most spectroscopic studies. Technical 
    issues, such as detector and telescope developments, are treated 
    separately from descriptions of specific missions and their results, 
    which contributed to the understanding of the UV sky. The missions are 
    compared in terms of their ``survey power'', a variable which combines sky 
    coverage and survey depth. I use the existing knowledge of UV sources to 
    predict views of the UV sky, which I then compare with those actually 
    detected. Finally, UV missions which will detect fainter 
    sources and will fly in the near future are described, and a wish list for 
    low-cost ventures, which could advance considerably our knowledge of the 
    UV sky is presented. 

       \end{abstract}

        \section{ Introduction } 
 
    Among all spectral bands, the ultraviolet has long been a neglected 
    region, in which we hardly have a good idea of how the sky looks like. 
    This is despite the fact that in the UV there is a distinct advantage of 
    small payloads: first, the sky is very dark, thus detection of faint 
    objects does not compete against an enhanced background (O'Connell 
    1987); second, the telescope construction techniques are very similar 
    (at least longward of $\sim$70 nm) to those used for optical astronomy. This 
    apparent neglect is likely to change in the foreseeable future. 
 
    Another factor, which presumably acted against the initiation of new sky 
    surveys in the UV, was the argument that not much can be learned about 
    the Universe from new UV sky surveys. After all, it was said, we can 
    infer about the appearance of the UV sky from a good data base in the 
    optical.  I shall show later that this is only approximately correct.
    The availability of a mechanism which allows one to predict the 
    appearance of the UV sky is a necessary ingredient in the detection of 
    outstanding sources, which may indicate the presence of new physical 
    phenomena. 
 
    The short astronomical history of our knowledge of the UV sky can be 
    divided into two eras, the first from the dawn of UV astronomy until the 
    flight of TD-1 and the second since the availability of the TD-1 all-sky 
    survey until today. Unfortunately, very little was accomplished in terms 
    of general sky surveys during this second era. We still await results 
    from a major, modern and sensitive UV all-sky survey, which hopefully 
    will be performed in the early 2000's. 
 
    Likewise, the UV domain may be divided into the ``regular'' ultraviolet, 
    that is, the segment from just shortward of the spectral region 
    observable from high ground-based observatories ($\sim$320 nm) to just 
    below the Lyman break at $\sim$90 nm, and the region from the Lyman 
    break to the fuzzy beginning of the X-ray domain. This short wavelength 
    limit may be arbitrarily defined as $\sim$6 nm ($\approx$200 eV). I 
    shall call the first segment ``the UV'' and the second the ``extreme UV'' 
    (EUV) bands. As shall be shown below, the observational techniques used 
    in the EUV band are more similar to those in X-ray astronomy, whereas 
    the UV is more like the optical. Sometimes, the UV region as defined
    here is separated into the near-UV and far-UV segments; the separation is
    at $\sim$200 nm.
 
    Although only few missions performed full sky surveys in the UV or EUV, 
    a large number of instruments scanned or imaged restricted sky areas. 
    Those provided partial, or very partial, information about the deeper UV 
    sky. By thorough examination of their results one may form an idea on 
    what could be expected from a full sky survey to a similar depth. A 
    review of UV imaging experiments and their results, updated to 1990, was 
    given by O'Connell (1991). A few UV imaging experiments were described 
    and compared by Rifatto \etal (1995), with the aim of understanding 
    galaxy UV emission. 

    In general, the sources of UV and EUV radiation can be either point or
    diffuse. A summary list of diverse sources, with the approximate order of
     importance in contribution, is shown below as Table 1,
    which is derived from the similar table of Gondhalekar (1990). A note of caution 
    should be mentioned here: the relevance of the importance of various
    contributions to the background is for broad-band measurements. In
    special circumstances, {\it e.g.,} for narrow-band or spectroscopic observations,
    line emission in the EUV domain may become the dominant contributor.
 
     \begin{table}[htb] 
    \begin{center} 
    \caption{Sources of UV and EUV radiation} 
     \begin{tabular}{ccc} 
    \hline  
    Source              & UV       & EUV     \\ \hline
    Direct starlight    & 1  & 1 \\ 
    Scattering off dust  & 2  & 3  \\ 
    Hot \& tepid ISM  & 6 & 2 \\ 
    H$_2$ fluorescence  & 3  & -- \\
    Galaxies & 4 & 4 \\
    AGNs  & 5 & 5 \\
    \hline 
    \end{tabular} 
    \end{center} 
     \end{table} 
    I shall not discuss  exotic mechanisms for the production of UV radiation 
    ({\it e.g.,} decays
    of exotic particles), except in a single case where a UV observation by HUT
    (see below) puts the most stringent limits on the mass of a heavy neutrino.
    The other sources shall be discussed below, in the appropriate context.

    It is clear that young stellar populations will emit
    copious amounts of UV radiation. To demonstrate this, I show in Figure 1
    how the peak of the spectral energy distribution changes with age in case
    of a starburst galaxy model of star formation. For almost 150 Myrs the peak stays 
    in the UV for the three different IMFs modeled here (Kroupa \etal 1993: 
    heavy solid line, Salpeter 1955: dotted line, and Miller \& Scalo 1979: filled 
    squares). All three models represent instantaneous bursts, where 50\% of 
    the stellar ejecta is recycled into stars, all include nebular emission,
    and in all 70\% of the Lyman continuum is absorbed by the ISM gas. The
    models were calculated with the codes of Fioc \& Rocca-Volmerange 
    (Leitherer \etal 1996). The steady UV peak for $\sim$150 Myrs is not due
    to Lyman $\alpha$ emission; a comparison with models which do not include
    nebular emission shows that this line can be important only during the few 
    first Myrs after the burst.

\begin{figure}[tbh]
\vspace{10cm}
\includegraphics{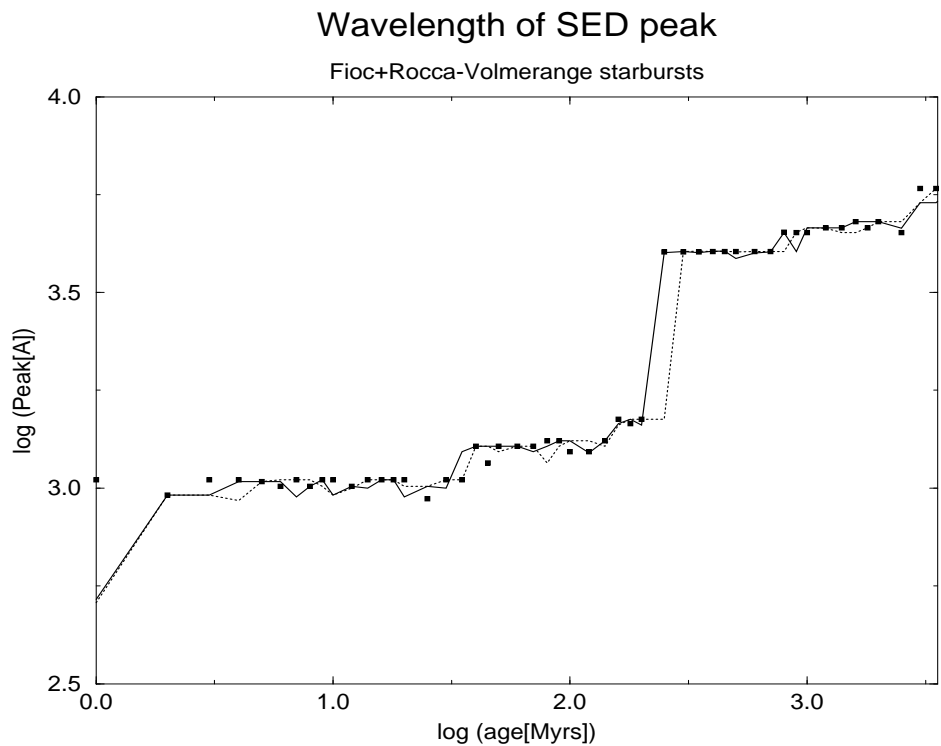}
\caption{Peak of spectral energy distribution of different starburst 
galaxy models {\it vs.} age. The models differ in the IMF used to calculate the
starburst. The dotted line represents the Salpeter IMF, the filled squares
are for the Miller \& Scalo IMF, and the solid line represents models calculated
with the Kroupa \etal (1993) IMF.}
\end{figure}
 
    Before embarking on a description of various missions and their results, 
    I discuss briefly the units used in UV astronomy. A useful description 
    of the brightness of astronomical sources is by ``monochromatic 
    magnitudes''. The monochromatic system is defined at wavelength $\lambda$
    (using the calibration of Vega by Hayes \& Latham 1975) as: 
\begin{equation} 
m_{\lambda}=-2.5 \, log(f_{\lambda})-21.175 
\end{equation} 
    where f$_{\lambda}$ is the source 
    flux density in erg/sec/cm$^2$/\AA\, at wavelength $\lambda$. A variant
    of the magnitude units is the AB system sometimes used in the UV. This is
    defined (Oke \& Gunn 1983) as:
\begin{equation}
AB=-2.5 \, log(f_{\nu})-48.60
\end{equation}
    The constant was chosen to make AB=V for objects with flat
    spectra. The units of $f_{\nu}$ are erg/sec/cm$^2$/Hz. The conversion
    between the two systems of magnitudes is wavelength-dependent. For
    $\lambda$=200 nm, which is representative of the UV domain, the
    conversion is:
\begin{equation}
m_{\lambda}=AB-2.26
\end{equation}

    Other 
    units, useful for describing background brightness, are ``photon units'' 
    (p.u. or c.u.=count units). These simply count the photon flux in a spectral 
    band, per cm$^2$, per steradian, and per \AA\,. At 200 nm, one count 
    unit equals 10$^{-11}$ erg/cm$^2$/sec/\AA\,/steradian, or   
    10$^{-13}$ W/m$^2$/nm/steradian, or 32.9 mag/arcsec$^2$. Of course, one can 
    alway use the ``flux units'' (erg/sec/cm$^2$/arcsec$^2$/\AA\,) directly, or 
    surface brightness units $\nu I_{\nu}$ (the units of I$_{\nu}$ are 
    erg/sec/cm$^2$/steradian/Hz). One unit of surface brightness equals 5 
    10$^7$ c.u.'s. 
 
    Waller {\it et al.} (1995) introduced the S$_{10}$ concept, used in the 
    visible part of the spectrum to describe the diffuse sky background, to 
    the UV domain. S$_{10}$ measures the number of 10th mag solar-type stars 
    per square degree which would produce the same flux level as observed 
    from the diffuse background. They quote for the region around 200 nm a 
    conversion of  
    \begin{equation}  
    1 \, S_{10} = 3.3 \times 10^{-21} erg/sec/cm^2/arcsec^2/\AA = 3.33 
    \times 10^{-10} c.u.  
    \end{equation}  
    As the UV background does not originate from Solar-type stars, it is not 
    clear what additional insight the S$_{10}$ unit introduces, except 
    when considering the influence of the zodiacal light. This is because 
    the evaluation of the zodiacal light contribution to the background 
    relies on a transformation from the visible band measurements by 
    Levasseur-Regourd \& Dumont (1980).  However, while the visible background 
    is  a manageable
    $\sim$200 S$_{10}$ (Roach \& Gordon 1973), the UV background is $\sim10^8$
    S$_{10}$, as will become clear below.
 
        \section{The Early Years. } 
 
        \subsection{ The "regular" UV } 
 
    The beginning of UV astronomy lies in the mid-1950s, with rocket flights 
    during which the skies were scanned by the free (unstabilized) flight of 
    the carrier vehicle. The first UV photometric data were obtained by 
    Byram {\it et al.} (1957), Kupperian \& Milligan (1957), and Boggess \& 
    Dunkelman (1959). In the early 1960s the first spectro-photometric 
    observations were obtained using three-axis stabilized rockets
    (Stecher \& Milligan 1962; Morton \& Spitzer 
    1966). However, rocket flights 
    offer only a limited amount of observing time, of order 5 minutes or 
    less, thus these surveys were limited in depth and sky coverage. 
 
    The first instrument on a satellite which was dedicated to UV astronomy 
    was a photometer, on a US Navy spacecraft launched in 1964 (Smith 1967). 
    One of NASA's first successful ``Observatories'' series of satellites was 
    OAO-2, with UV cameras from the Smithsonian Observatory and a 
    medium-dispersion spectrograph from GSFC, launched on 7 December 1968 
    (Davis {\it et al.} 1972, Code \etal 1970). The CELESCOPE imaging 
    experiment, with its four 30 cm diameter telescopes, imaged 2$^{\circ}$ 
    wide fields with relatively low angular resolution. Its results, 
    consisting of photometric measurements of some 5,068 stars in four 
    spectral bands, make up the CELESCOPE catalogue (Davis \etal 1973). 
 
    Even manned space flights were harnessed into providing UV astronomical 
    information; John Glenn was given a hand-held 35 mm camera with an 
    objective prism to record the UV spectra of stars, but nobody checked 
    whether the window of Freedom-7 was UV-transparent (it was not, and no 
    images were obtained; Boggess \& Wilson 1987). Later, manned flights 
    provided valuable UV data 
    from hand-held cameras on Gemini flights (Henize {\it et al.} 1975) and 
    from an automated Moon-based telescope (Carruthers 1973). A similar
    attempt at UV astrophysics from a manned space platform was done
    on 19-24 December 1973, when the space observatory ORION-2 was
    operated from the Soyuz-13 spacecraft (Gurzadyan \etal 1985). The
    ORION-2 telescope was a 22 cm Cassegrain with an objective prism,
    and the spectra of $\sim$900 stars were recorded on film.
 
    Some information on the diffuse UV background between 135 and 148 nm was 
    published by Hayakawa {\it et al.} (1969). The measurements were done 
    during a rocket flight on 7 March 1967 and the experiment, along with 
    three-band photometry of six bright stars, as described by Yamashita 
    (1968). Approximately half the sky was mapped on 1 March 1970 in a UV 
    band from 142.5 to 164 nm and with coarse angular resolution of about 
    10$^{\circ}$ by a rocket-borne Geiger counter which had an effective 
    collecting area of 0.62 cm$^2$. The experiment and its results were 
    described by Henry {\it et al.} (1977). The results were compared with a 
    model of the UV sky calculated by Henry (1977, see below). 
 
    The UV spectral range is blessed with very dark skies (O'Connell 1987), 
    which permits achieving a good S/N ratio on faint sources with relatively 
    small optics, provided one has a noiseless detector and keeps away from 
    sources of background radiation. One well-known source of 
    background emission is the geocoronal Lyman $\alpha$ (Ly$\alpha$), 
    resonant scattering of Solar photons off hydrogen atoms in the halo 
    around the Earth. A beautiful image showing the entire Earth with its 
    Ly$\alpha$ halo, as well as the tropical UV airglow bands and the 
    auroral ovals around the poles, has been obtained by the S201 experiment 
    operating from the Moon during the Apollo 16 mission (Page {\it et al.} 
    1982, Fig. 4b). 

    The auroral emission is present essentially everywhere 
    around the Earth, but its intensity varies with geomagnetic latitude 
    (Meier 1991). The Ly$\alpha$ emission, and the aurora and airglow, are 
    the two most troublesome backgrounds influencing orbiting UV experiments 
    at low and intermediate altitudes. The
    OI line emission at 130.2 and 135.6 nm band, and the N$_2$ Lyman-Birge-Hopfield
    bands in the 140-180 nm region, are restricted to the upper atmosphere
    and affect only observations done at low orbital altitudes, or at
    low incidence angles to the Earth limb. The OI lines, in particular,
    form at altitudes of 250-300 km (Leinert \etal 1998). Special LEO missions may,
    therefore, require special tailoring of mirror coatings and of
    filters to suppress this background. Note that Ly$\alpha$ resonant 
    scattering is produced also off interplanetary hydrogen as well as off 
    interstellar H$^0$ atoms passing through interplanetary space, thus no 
    location in the Solar System is really free of this emission. 
 
        \subsection{ The extreme UV } 
 
    In parallel with the development of the UV astronomy field, first steps 
    were taken to study the EUV sky. The EUV range is hampered by the 
    opacity of the interstellar medium (ISM). From 91.2 nm shortward to 
    about 10 nm the opacity is high, because of the photo-electric 
    cross-section of hydrogen, and to a lesser extent of neutral helium 
    (below 50.4 nm) and singly ionized helium (below 22.8 nm). The opacity 
    of the ISM limits the detectable range to about 100 pc., with very few 
    exceptions. The objects expected to be detectable in the EUV were hot 
    early-type stars, hot white dwarfs, coronae of late-type stars, and 
    bright non-thermal sources some of them possibly extragalactic. Note
    that some of the EUV emission from early-type stars was much more
    intense than expected from model atmospheres ({\it e.g.,}
    $\epsilon$ CMa; Vallerga \etal 1993, Wilkinson \etal 1996). Also, 
    not many {\bf very} hot
    young stars were expected to be detectable because of the significant
    intervening hydrogen column density to such objects.
 
    The first studies in the EUV range were mostly, as in the UV domain, by 
    rocket-flown instruments (Henry {\it et al.} 1975 a, b, c), which 
    measured a few very bright sources and established calibrators. The 
    earliest observations below Ly$\alpha$ were by Belyaev {\it et al.} 
    (1971) with Geiger counters on the Venera 5 and 6 spacecraft. These 
    attempts were done with large field of view detectors and concentrated 
    on the detection of the EUV background. The EUV background was also 
    characterized (Kumar {\it et al.} 1974; Bowyer {\it et al.} 1977, 1981). 
    The culmination of this work was the EUV instrument flown on the 
    Apollo-Soyuz mission in 1975, when four EUV point sources were 
    discovered (Lampton {\it et al.} 1976, Margon {\it et al.} 1976, Haisch 
    {\it et al.} 1977, Margon {\it et al.} 1978). At the completion of 
    these, first, preliminary pilot surveys, NASA initiated the EUVE survey,
    described in section 5.3. 
 
    The Voyager spacecraft explored the EUV sky with their Ultraviolet 
    Spectrometers (UVS: Sandel, Shemansky \& Broadfoot 1979). For a number 
    years, the UVS on the two Voyager spacecraft were the most distant 
    astronomical observatories in operation (Holberg 1990, 1991). The
    operation of the UVS instruments is scheduled to end in 1998, in order to
    conserve power for other Voyager instruments as the radioisotope power
    geenrators degrade. A look 
    onto the EUV sky, as side-benefit of an X-ray mission, was also offered 
    by the ESA mission of EXOSAT. This sky survey was performed from 0.6 to 
    40 nm and covered the short wavelength end of the EUV range. The EXOSAT 
    results were summarized by White (1991).

        \section{Technical issues } 
 
    Before continuing with the description of UV missions in ``modern'' 
    times, it is necessary to discuss technical issues involved in UV 
    observations. These have to do with both telescope construction and 
    detector development. Both aspects are driven not so much by scientific 
    aspects as by military usage of the UV, mainly for targeting missile 
    launches and re-entry bodies. 
 
        \subsection{ Optics } 
 
    In the telescope construction section we recognize that down to about 
    115 or 105 nm (and in some cases even below that)
    ``regular'' telescope construction techniques apply. These 
    imply parabolic-hyperbolic mirror combinations (Cassegrain, 
    Ritchey-Chr\'{e}tien, or even Schmidt configurations) made of glasses, 
    ceramic materials, or light-weight metals. The primary difference from 
    ground-based telescopes is that for space devices light-weightedness is 
    a necessity. Therefore, space mirrors will usually be light-weighted by 
    machining the blanks to reduce the mass as much as possible. 
    Light-weighting to 40\% of a solid blank is customary, although reports 
    were published of light-weighted mirrors to 20\% or lower values. 
 
    The requirements for a reflecting coating for UV wavelengths above 115 
    nm are fulfilled by 
    aluminum overcoated by MgF$_2$. Another 10 nm to shorter 
    wavelengths can be gained if the coating is LiF. Note that both 
    materials are hygroscopic (LiF more than MgF$_2$) and the mirrors have 
    to be maintained in a controlled atmosphere until deployed in space. For 
    UV observations below 115 nm but above $\sim$50 nm the coating of choice 
    has until recently been osmium or iridium. These materials are inert at 
    ground-level atmospheric conditions and offer reasonable 
    reflectivities of $\sim$20\%. Recent missions (second flight of HUT, 
    second flight of ORFEUS, etc.) used SiC mirror coatings. These allow
    high-efficiency normal-incidence reflections for $\lambda\geq$60 nm
    (Keshi-Kuha \etal 1997). This mirror option was chosen also for the FUSE
    mission (Kennedy \etal 1996).
 
    One important issue with UV optics, as well as with EUV, is optics 
    contamination. Not only must dust particles be kept away, but also 
    molecular contaminants. These are harder to drive off, because some have 
    high molecular weights and will not evaporate fully during outgassing 
    procedures. In space, under the influence of high-energy particles and 
    UV photons, these substances change chemical form, may polymerize, and can
    form mono-layers over the mirrors with very high optical depths in 
    the UV (Noter {\it et al.} 1993). The significant reduction of the UV 
    throughput of HST's WFPC-1 by molecular contaminants, polymerized under 
    the diffuse UV radiation from the Earth, was confirmed by MacKenty {\it 
    et al.} (1995). It is possible that a similar phenomenon was partially
    responsible
    for the sensitivity decrease of HUT in orbit (for a discussion of
    the HUT sensitivity see Davidsen \etal 1992).
 
    Observations in the EUV range cannot rely (with one exception) on 
    normal-incidence optics. For this spectral region the geometry of the 
    telescope mirrors follows that used in the X-ray domain, the Woltjer 
    pairs, where the reflection is by grazing incidence. The one exception 
    to this rule is for narrow-band observations where the mirror acts as 
    filter, by being coated with multi-layers of different metals. To 
    improve the efficiency, the telescope architecture normally used in 
    these cases is with the detector at the prime focus of a single 
    parabolic mirror. This technique was used for the ALEXIS telescopes and 
    was planned for the first generation design of the EUVITA telescopes for 
    the Spectrum X-$\gamma$ (SRG) spacecraft (described by Courvoisier {\it 
    et al.} 1993; the option to orbit this configuration has since been 
    abandoned). 
 
    In principle, multi-layer coatings can yield a total in-band reflectivity 
    of about 40\%, with a bandwidth of about 10\% of the peak band 
    wavelength (Roussel-Dupr\'{e} \& Ameduri 1993). In practice, it is 
    extremely difficult to produce multi-layer coatings on mirror substrates 
    larger than about 20 cm which are uniform, stable, and can survive the 
    launch environmental stresses. 
 
        \subsection{ Detectors } 
 
    The detector issue has been driven for many years by requirements of as 
    high as possible quantum efficiency together with as good a resolution 
    as possible. The first detectors used in UV and EUV astronomy were 
    photomultipliers with cathodes sensitive to the spectral bands to be 
    observed, or were just Geiger counters. TD-1 (see below) achieved 
    spatial resolution by scanning the photomultiplier aperture on the sky. 
 
    Most modern UV imaging detectors are intensifiers or image converters, 
    which use a UV-sensitive cathode as their main active ingredient. In 
    most experiments the images were recorded on film. This was processed on 
    the ground after the observation and was usually scanned to convert the 
    information into digital pictures. Among the experiments which used this 
    mode of image recording we count SCAP-2000, FOCA, WF-UVCAM, FAUST
    (first flight only), and 
    UIT. These shall be discussed in detail below. At this point, it is 
    worthwhile to remark that the main limitations of this method of image 
    recording are (i) the limited and non-linear dynamic range, and (ii) 
    possible film defects, that can be detected only after processing. 
 
    Carruthers (1973) developed a UV electronographic camera, patterned 
    after visible-light ECAMs. This camera extracted photo-electrons from a 
    cathode by UV light and accelerated them over $\sim$1000V potential 
    difference. The electron trajectories were confined by the strong 
    magnetic field, parallel to the optical axis, of a permanent magnet 
    wrapped around the camera. The image was created by the energetic 
    electrons onto special electronographic emulsion placed at the focal 
    plane. The readout was accomplished by micro-densitometry of the exposed 
    and developed emulsion. To achieve a better dynamic range, the same 
    scene was imaged with different exposure times. Modern versions of 
    Carruthers' cameras employ electron-bombarded CCDs (EBCCDs) in place of 
    the electronographic emulsion; each accelerated electron creates many 
    electron-hole pairs in the CCD, allowing high S/N detection of 
    individual photons. The obvious advantage of Carruther's design is
    that the quantum efficiency of the opaque cathodes he uses is much
    higher than that of the semi-transparent cathodes used in other
    designs. The disadvantage is the need of an extended, strong, and 
    uniform magnetic field along the axis of the camera. This limits the
    size of the cameras built by his group to fairly small apertures.
 
    It is also possible to use CCDs for UV observations, as done in the 
    WFPC-1 and WFPC-2 of the HST. Even better UV performance is achieved by 
    modern CCDs, with thinned, back-illuminated chips which have 
    fluorescing, anti-reflection coatings. A comparison of CCD and other
    detectors, with specific application to low light level observations as 
    encountered in astronomical situations, was done by Vallerga \& Lampton 
    (1987). While the authors concluded that the CCD and MCP (see below) 
    devices are equivalent 
    in terms of required number of incident photons to reach a set S/N, the 
    disadvantage inherent in the cryo-cooling required for most CCDs is 
    evident. In addition, MCP-based detectors with electronic readout
    offer readily a photon-counting option.
 
    With the advent of multi-channel plates (MCPs) it became possible to 
    build position-sensitive detectors in the UV and EUV where photons could 
    be ``confined'' to paths within individual channels, while being 
    accelerated and multiplied. With three-stack MCPs, gains of 10$^6$ to 
    10$^7$ can be achieved. Examples of such detectors are those of FAUST 
    (in its second Shuttle flight), and EUVE, where the readout is done by a 
    position-sensitive anode (wedge-and-strip). The design of TAUVEX (see 
    below) is based on similar detectors. A review of readout methods for 
    photon-counting MCP-based detectors was presented by Lampton (1987). 
 
    At least three other forms of readout are available for MCP-based 
    detectors. It is possible to have a phosphor output activated by the 
    electron cloud emerging from the last MCP of the stack, which is 
    optically-coupled to a regular, fast-readout CCD. Special fast phosphors, 
    with fast analyzing circuitry and hardwired, or transputer-based 
    centroiding algorithms, yield sub-pixel resolution of the center of each 
    electron cloud. The resultant detector module is thus high-resolution, 
    fast, and photon-counting. Similar detectors are operating on the MSX 
    UVISI and on the future UV/optical monitor of XMM (see below). Recent
    results with this configuration report resolution below 6 $\mu$m,
    resolving individual MCP pores, while supporting high count
    rates (Vallerga \etal 1997).
 
    A different readout mode employs a resistive anode, with outputs at 
    opposite locations at the the anode edges and fast timing circuits to 
    measure the propagation delay of the charge cloud to each of the 
    outputs. The centroid of the charge cloud is obtained by differencing 
    the time delays in the same direction. The new detector of FOCA (see 
    below) operates in this mode. Finally, another position-sensitive anode 
    uses crossed pairs of helical delay lines behind the last MCP. The 
    readout is conducted by measuring time differences between the charge 
    pulses received at the two ends of a wire. Reported performance of 
    helical delay line readouts appears to offer very high data rates (up to 
    MHz counts) and very high resolution (about 20 $\mu$m, Siegmund {\it 
    et al.} 1992). This, apparently, is the readout of choice for the 
    GALEX UV sky survey mission.
 
    The HST instrument Space Telescope Imaging Spectrometer (STIS),   
    installed during the refurbishing flight of 1997, includes one CCD 
    detector for $\lambda>200$ nm and two Multi-Anode Microchannel Array 
    (MAMA) cameras for the 115-170 and 165-310 nm bands. The MAMAs are 
    essentially MCP intensifiers with CsI or Cs$_2$Te cathodes, where the 
    location of the electron clouds is achieved by centroiding the readout 
    of crossed micro-strip anodes. 

    A brief discussion of current detector 
    technology for UV astronomy was presented by Ulmer \etal (1995), and a more 
    extensive discussion is by Joseph (1995). 

    Two kinds of promising detectors have not yet been used in UV astronomy.
    One is the low-pressure multistep gaseous electron multiplier, which is
    used in high-energy physics for detecting \v{C}erenkov photons in ring-imaging 
    detectors ({\it e.g.,}
    Chechik \& Breskin 1988). Such detectors can reach very high UV quantum
    efficiencies, up to 60\% depending on the type of gas filling the detector
    (A. Breskin, private communication), but have
    never been packaged for space applications. The other type of detector is
    superconducting tunnel junctions packaged in an array configuration ({\it e.g.,}
    Perryman \etal 1994), which may prove to be the first energy-sensitive
    imaging array for UV applications. This will be discussed below, near the end
    of this paper.

    New developments indicate that it may be possible to fabricate array 
    devices similar to CCDs which will have good UV sensitivitiy in the UV coupled
    with sharp response cutoffs near the visible by using wide bandgap III-IV
    semiconductor materials, such as GaN and other nitrides ({\it e.g.,}
    Kung \etal 1996).
 
        \section{ The TD-1 Era } 

    The ``middle-ages'' of UV and EUV astronomy witnessed a number of high-atmosphere
    or space missions, which shall be discussed below. In order to put these in proper
    perspective, I present in Table 2 a summary of vital statistics for these missions.
    The spectral resolution is given in $\frac{\lambda}{\Delta\lambda}$ in the
    usual spectroscopic convention, where $\Delta\lambda$ is
    the typical width of the spectral bands used by each instrument. Only TD-1,
    IUE, and EUVE had spectroscopic capabilities. I include
    the instantaneous field-of-view of each instrument in column 2.  The table 
    is arranged in chronological
    order of missions and concentrates the three EUV surveys at its bottom. The
entry for EUVE includes the approximate spectral resolution of the imagers as well as
that of the spectroscopic telescopes.
   
    \begin{table}[htb]  
    \begin{center}  
    \caption{ UV and EUV survey missions}  
    \begin{tabular}{lllllllll}  
    \hline  
   Mission & Inst.      & Apert. & Spectral & Spatial  & Spectral   & Launch & End of & Responsible \\
    name &   FOV (deg.) & cm.      & range (nm)  & resol'n & resol'n & year   & mission & agency \\
    \hline  
    TD-1   & 0.25          & 27.5  & 157-274 & 2'        & 6          & 1972   & 1974  &  ESRO \\
    S201   & 20            & 7     & 125-160 & 3'        & NA         & 1972   & 1972  & NASA (NRL)  \\  
    FUVCAM & 11-20         & 10    & 123-200 & 3'        & 3          & 1978   & 1991  & NASA (NRL)  \\  
    IUE    & 4 10$^{-3}$   & 45    & 110-320 & NA        & 285        & 1978   & 1996  & NASA, ESA, SERC \\  
 SCAP-2000 & 6             & 13    & 190-210 & 2'        & NA         & 1979   & 1990  & CNES, CNRS, FNRS \\  
  GSFC CAM & 11.4          & 31    & 140-262 & 1'        & 2          & 1979   & 1980  & NASA (GSFC) \\ 
  WF-UVC   & 66            & 20    & 125-280 & 5'        & 1          & 1983   & 1983  & CNES (LAS) \\  
    GLAZAR & 1.3           & 40    & 150-180 &   10"-40" & NA         & 1986   & 1991  & Russia, Armenia \\  
    GUV    & 4             & 17    & 130-164 & 12'       & NA         & 1987   & 1987  & Japan (ISAS) \\  
    UIT    & 0.67          & 38    & 125-290 &  3"       & 2          & 1990   & 1995  & NASA (GSFC) \\
    FOCA   & 1.5, 2.3      & 39    & 190-210 & 10"-20"   & NA         & 1991   &  --   & CNES, CNRS, FNRS \\  
    FAUST  & 8             & 16.1  & 140-180 & 3'.5      & NA         & 1992   & 1992  & NASA, UCB, CNES \\     
    \hline 
    WFC    & 5             & 57.6  & 17-210 eV & 2'      & 9, 120 eV  & 1990   & --    & SERC (Leicester) \\  
    EUVE   & 5             & 40    & 10-60    & 1'       & 1, 200     & 1992   & --    & NASA (Berkeley) \\
    ALEXIS & 33            & 10    & 13.3-18.8 & 15'     & 10         & 1993   & --    & NASA (LANL) \\  
    \hline  
    \end{tabular}  
    \end{center}  
 
       \end{table}

        \subsection{ TD-1 } 
 
    The advent of modern UV astronomy can be marked as the first UV all-sky 
    survey to a reasonable depth and with a reasonable angular resolution, 
    by the TD-1 satellite. As one of the first satellites launched on 12 
    March 1972 by the European Space Research Agency (ESRO, later known as 
    ESA), the UV experiment was known as S2/68 (the second scientific 
    satellite of year 1968). The UV experiment was a collaboration of the 
    British and Belgian scientists, and is described by Boksenberg {\it et 
    al.} (1973). It consisted of an f/3.5 telescope with a 27.5 cm diameter 
    primary mirror, feeding a spectrometer for the 130-255 nm region and a 
    photometer with a single broad band centered at 275 nm. The entrance 
    aperture of the spectrometer was 11'.8$\times$17' and the photometer 
    aperture was 1'.7$\times$17'. The calibration of TD-1 was described also 
    by Humphreys \etal (1976). 
 
    Already during its operational period the TD-1 observations were used to 
    search for outstanding UV-bright objects (Carnochan {\it et al.} 1975). 
    The results, in the form of an all-sky catalog of UV sources, were 
    published by Thompson {\it et al.} (1978). In order to produce the
    catalog, the spectroscopic data were binned into three photometric
    channels, each 33 nm wide: 135-175 nm, 175-215 nm, and 215-255 nm.
    These, and the purely photometric 274 nm (31 nm wide) channel, are
    the four TD-1 bands. The ESA-TD1 catalog contains 
    31,215 stars measured with S/N$>$10 in all four TD-1 bands. An 
    unpublished version, with lower S/N restrictions, has 58,012 objects 
    (Landsman 1984). 
 
    The results of TD-1 were discussed by many and it appears that the TD-1 
    sensitivity limit was not uniform across the sky. This was because of 
    the variable number of scan repetitions of the same source (higher at 
    high ecliptic latitude), and also probably because of variations of UV 
    background, mainly in the 156.5 nm band due to geocoronal Ly$\alpha$ 
    emission (Morgan {\it et al.} 1976). Gondhalekar {\it et al.} (1985) 
    discussed the TD-1 results in the context of the galactic UV 
    interstellar radiation field emission. They mentioned that TD-1 is 
    probably not linear for fluxes fainter than 10$^{-12}$ 
    erg/s/cm$^2$/\AA\,. Henry (1991) noted that the dark current of TD-1 
    varied with time, another possible reason for non-linearity at low 
    signal levels.

\begin{figure}[tbh]
\vspace{9cm}
\includegraphics{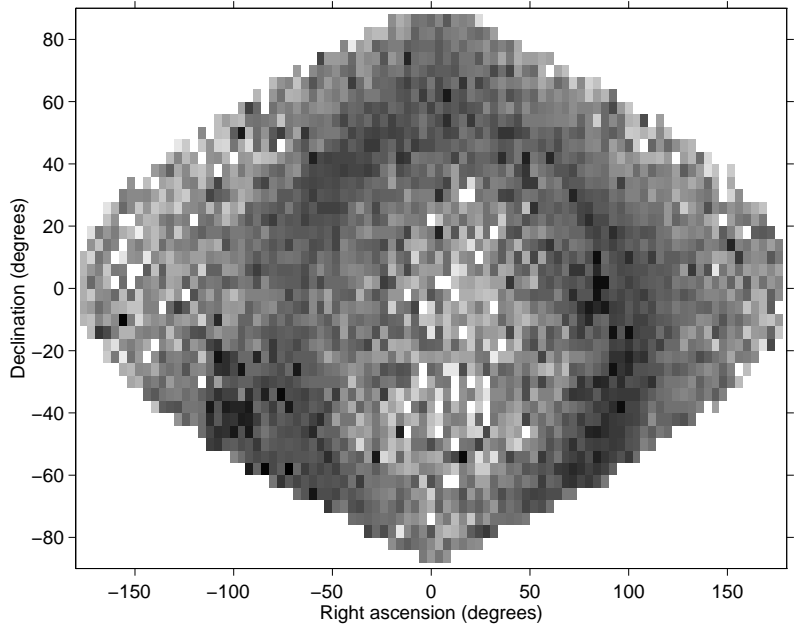}
\caption{The TD-1 sky at 156.5 nm. The 31,215 sources of the published catalog 
were binned (in flux) into ``fat'' pixels to provide a fuzzy view of the entire sky. 
The intensity is shown logarithmically, to stetch the grey scale, with brighter
regions represented as darker pixels.}
\end{figure}
 
    I show in Figures 2 and 3 the UV sky seen by TD-1. This was obtained by binning
    the flux from the stars in the published catalog in boxes of
    4$^{\circ}\times4^{\circ}$, and displaying the resultant average surface brightness
    with a logarithmic scale. The view is in celestial coordinates; the Milky
    Way is the darker (brighter) ``bullseye'' ring centered on both figures. The high
    flux area in the Milky Way, visible on the upper right side of the image,
    is the Orion region. The dark region in the diametrically opposite 
    direction is produced by UV-bright stars in Scorpius-Sagittarius.
    Nearby UV-bright stars produce isolated black fat pixels at random locations
    in the figures.

\begin{figure}[tbh]
\vspace{9cm}
\includegraphics{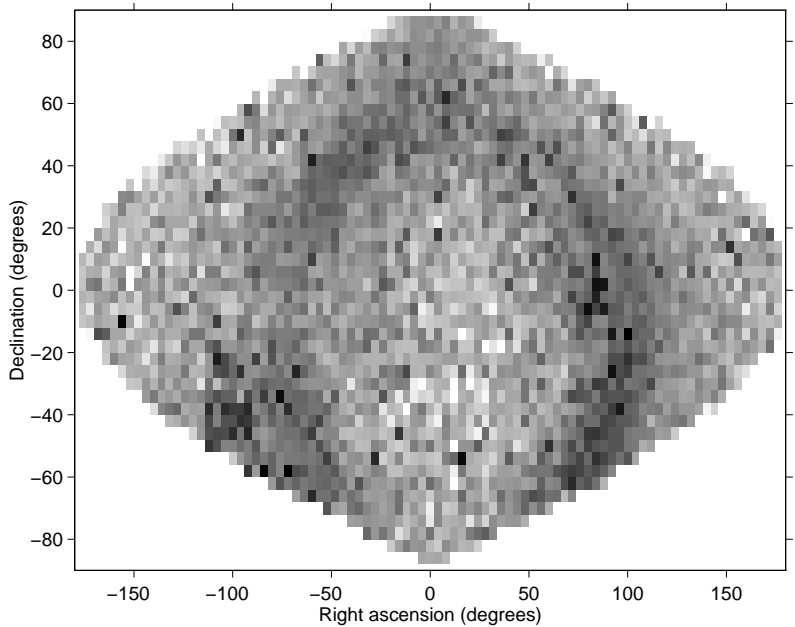}
\caption{The TD-1 sky at 274 nm. The format of the display is identical to 
that in Fig. 2.}
\end{figure}

    Based on the TD-1 results, Henry \etal (1988) produced an ``Atlas of the 
    Ultraviolet Sky'', which combines plots of the visible sky and of the
    corresponding 156.6 nm view. The faintest objects shown in these plots
    are at 10$^{-12}$ erg cm$^{-2}$ s$^{-1}$ \AA\,$^{-1}$, but among the
    objects fainter than five times this value, some have had the UV
    brightness calculated from an optical-to-UV transformation.
 
    The all-sky coverage and the relative depth 
    of the TD-1 S2/68 experiment have set its results as a benchmark 
    against which all other sky surveys are and will be measured. The TD-1 
    catalogs, both the published S/N$>$10 version and the lower S/N 
    unpublished version, are possible mines of interesting objects. An 
    example of the lasting value of the TD-1 survey is the report by 
    Landsman {\it et al.} (1996), who found two cases of white dwarf 
    components in binaries with F-star primaries. They selected sources with 
    TD-1 156.5 nm UV excesses identified in the catalog as late-type stars 
    and followed them up with IUE. In two cases (56 Peg and HR3643), UV 
    signatures of white dwarf stars were detected. 
 
    Despite this long-lasting value, the TD-1 catalog is limited in the sky 
    view it presents. The TD-1 survey is as shallow in the UV as the HD 
    catalog is in the optical; the sky knowledge it offers through the total 
    number of sources is equivalent to that of the visible sky as was known 
    in the early stages of astronomical photography, about 100 years ago 
    (O'Connell 1992)! After TD-1, the various UV and EUV efforts can be 
    characterized as either imagers or spectrometers. Among the imagers, 
    some were orbiters and others were on short-duration flights (rockets, 
    balloons, high-altitude planes). I shall list the various missions 
    below, concentrating almost exclusively on imagers. 
 
        \subsection{ANS } 
 
    This mission was launched on 30 August 1974 and was described by Van 
    Duinen {\it et al.} (1975), by Wesselius {\it et al.} (1982) and by de 
    Boer (1982). The instrument consisted of a 22.5 cm diameter telescope, 
    which focuses the light through a 2'.5$\times$2'.5 slit onto a 
    spectrometer with fixed slits in its focal surface. Each location is 
    thus sampled in five spectral bands, defined by the sizes and locations 
    of these exit slits. The spectral coverage of ANS was from 150 nm to 325 
    nm and each band was about 15-20 nm wide. The maximal sensitivity of 
    ANS, determined by the instrument design, was at 220 nm. 
 
    ANS did not perform a sky survey, but observed pre-selected targets. 
    Wesselius {\it et al.} (1982) reported that only 3,573 objects out of 
    the more than 5,000 observed by ANS were actually retained in the final 
    catalog of the mission. These were objects with S/N$>4$ in one band, or 
    S/N$>3$ in at least three of the ANS bands. Measurements of 13 
    elliptical galaxies were reported by de Boer (1982); in many cases only 
    upper limits could be presented, because of the low reflecting 
    efficiency of the grating, away from its blaze angle. 
 
    One important usage of the ANS data set was to derive UV-to-optical 
    color indices from observations of 182 main sequence stars and 56 giants 
    and supergiants (Wesselius {\it et al.} 1980). Note that the derivation 
    of color runs only down to G0 stars; later-type stars were apparently 
    too faint for ANS to register. This is a characteristic of most UV 
    surveys; the late-type stars are under-represented, or absent 
    altogether. 
 
        \subsection{ The IUE observatory } 
 
    Undoubtely, the greatest success ever of any orbiting astronomical 
    instrument lies with the IUE observatory. Launched on 26 January 1978 
    for a nominal three years' mission, the observatory operated for 18 
    years yielding more than 100,000 spectra of nearly 9600 diverse 
    astronomical targets. IUE operations closed down on 30 September 1996. 
    The experiment consists of a 45 cm 
    diameter Cassegrain telescope feeding two alternative spectrometers, one 
    for the range 110-190 nm and the other for 180-320 nm. Two different 
    dispersions were available, low (R=300) and high (R=2 10$^4$, with an 
    echelle arrangement). The sensitivity of IUE, for low dispersion 
    operations, was aproximately m$_{150}$=15. An early description of IUE 
    can be found in Boggess \etal (1978). 
 
    During the unexpectedly long operation period of IUE, its NASA operators 
    at GSFC and ESA personnel at VILSPA devised work-around methods to do 
    with less than the minimal number of gyros, and to operate despite 
    unexpectedly high levels of straylight when a piece of thermal blanket 
    or reflecting tape fluttered in front of the telescope aperture.  
 
    Although the IUE data set does not represent a uniform survey of the 
    sky, the large variety of objects observed by it offers unique opportunities 
    to derive ``average'' properties of populations. This has been used by 
    many ({\it e.g.,} Fanelli {\it et al.} 1987) to derive UV-to-optical color 
    indices for various spectral types and luminosity classes. These are 
    later used to derive transformations, to create models of the UV sky (see 
    below), or to determine the level of the diffuse UV background. 
    Observations of galaxies are used to determine average UV spectra of 
    irregular, spiral, and elliptical galaxies, and of galactic bulges 
    (Ellis \etal 1982, Burstein {\it et al.} 1988, Kinney {\it et al.} 
    1993, Storchi-Bergmann 
    {\it et al.} 1994), important for the derivation of k-corrections. 
 
    The IUE data bank represents a valuable archival resource, even more so 
    after the final reprocessing of all the low-dispersion spectra into the 
    final Uniform Low-Dispersion Archive (ULDA) will be complete. To help 
    the logical usage of the ULDA, atlases of UV spectra of selected types 
    of objects, based on IUE data were published ({\it i.e.,} Longo \& 
    Capaccioli 1992). Apart from the special-purpose atlases, note those 
    dedicated to the classification of UV stars by ESA and by NASA   
    (Heck \etal 1984; Wu \etal 1991). The last distributed version of 
    ULDA (V4.0), including 
    54,247 spectra obtained until 31 December 1991, has been installed at 27 
    regional or national centers. The final version of the archives 
    contains 104,471 spectra reprocessed with the most up-to-date calibration
    and includes echelle spectra binned to the resolution of the low
    dispersion observations.
 
        \subsection{ASTRON space station} 
 
    The ASTRON space station was launched by the USSR on 23 March 1983. It 
    was built on a Venera-type platform and included a 
    3$^{\circ}\times3^{\circ}$ (FWHM) proportional counter for observations 
    in the 2-25 keV X-ray regions and a 80 cm telescope for UV astronomy. 
    ASTRON operated from a highly elliptical orbit ($\sim$2000 km $\times 
    \sim$ 200,000 km, four-day duration) until June 1989. Some results on galaxy 
    photometry in 2.8 nm wide bands, from 160 to 350 nm, were published by 
    Merkulova \etal (1990). Spectroscopy of SN1987A, and of flares on the 
    red dwarf EV Lac, were reported by Liubimkov (1990), by Burnasheva \etal 
    (1989), and by Katsova \& Livshits (1989). An interesting and unique 
    feature was the ability to perform fast UV spectro-photometry, with a 
    time resolution of 0.61 sec (Katsova \& Livshits 1989). 
 
    The UV telescope on the ASTRON station did not operate in a survey mode, 
    and the number of objects observed by it was probably very restricted. 
 
        \subsection{ S201 } 
 
    The NRL experiment S201 was described by Page {\it et al.} (1982), where
    the revised list of sources is given. A 
    summary of results was given by Carruthers \& Page (1984c). It consisted 
    of an electrographic Schmidt camera, which operated automatically on the 
    Moon during the Apollo 16 mission in April 1972. The camera had a field 
    of view of 20$^{\circ}$ and an angular resolution of about 3'. The 
    limiting magnitude of the longest exposures, about 30 minutes long, was 
    m$_{UV}$=11 (typically m$_{UV}$=10) in the spectral band 125-160 nm. The camera 
    did not compensate for the rotation of the Moon. This caused trailing of 
    the long exposure images, by 0'.54 per minute. 
 
    The S201 experiment obtained the first UV image of the LMC (Page \& 
    Carruthers 1981) and demonstrated the difference between its optical 
    appearance, dominated by old, evolved stars, and the UV where hot, young 
    stars dominate. In total, ten fields each 20$^{\circ}$ in diameter were 
    observed, thus the experiment covered $\sim$7\% of the sky. The results 
    of the S201 experiment were discussed in a series of papers (Carruthers 
    \& Page 1983, 1984a, 1984b). A discussion of the UV properties of 
    nebulae in Cygnus was published by Carruthers \& Page (1976).

        \subsection{ GUV } 
 
    The GUV experiment, flown on 21 February 1987 on sub-orbital flight 
    with the S520-8 
    rocket, was described by Onaka {\it et al.} (1989). It consisted of two 
    17 cm diameter Ritchey-Chr\'{e}tien telescopes, which imaged fields 
    4$^{\circ}$ in diameter. The telescopes were of f/3.2 F=36 cm design, 
    and the detectors used were CsI cathodes, tandem MCPs, and resistive 
    anode readouts. The cathode response, combined with the transmission of 
    the BaF$_2$ windows, 
    defined a spectral band centered at 142 nm  with a FWHM of 22 nm. The 
    two sky field positions were offset by about 3 degrees and the final 
    angular
    resolution, 16'x8'.2, was determined by the pixel size (2'.4), the 
    quality of the optics, and the stability of the platform in the pointing 
    phase. 
 
    The GUV experiment produced rather shallow stellar photometric data on 
    general sky regions during the 133 second spin-stabilized ascent, in 
    which 48 stars were detected, and deeper one-band photometry of the 
    Virgo cluster during the pointed phase, which lasted 181 seconds. The 
    limiting magnitude of the GUV experiment was approximately m$_{156}$=14.6. 
    Unfortunately, the recovery of the S520-8 payload was not successful and 
    the payload sank at sea, preventing the possibility of post-flight 
    recalibration. 
 
    The GUV observations were re-analyzed by Kodaira {\it et al.} (1990), 
    and more than 40 galaxies were detected in the Virgo cluster. The 
    authors found correlations of UV emission with HI flux and with FIR 
    emission for spiral galaxies, and with X-ray and radio emission for 
    elliptical galaxies. 
 
        \subsection{ SCAP 2000 } 
 
    This is a very interesting collaboration between the Observatoire de 
    G\'{e}n\`{e}ve and the Laboratoire d'Astrophysique Spatiale du CNRS of Marseille, 
    running over more than two decades. The collaboration (supported by CNES 
    and CNRS in France, and by FNRS in Switzerland) produced a stabilized 
    balloon gondola (Huguenin \& Magnan 1978) carrying a telescope tuned for 
    imaging observations in the UV. The experiment was described by Laget 
    (1980), by Donas {\it et al.} (1981), and by Milliard {\it et al.} 
    (1983). It consists of a 13 cm Schmidt-Cassegrain reflector with a field 
    of view of 6$^{\circ}$ and an effective collecting area of $\approx$95 
    cm$^2$.
 
    The detection was accomplished by an image converter-intensifier, which 
    together with the mirror coatings and absorption of the few mbar 
    atmosphere at the balloon altitude defined a bandpass centered at 
    $\sim$200 nm and $\sim$15 nm wide, and was acheived mainly by multi-layer 
    coatings of the primary and secondary mirrors. The bandpass in identical 
    to that of the newer experiment FOCA (Milliard {\it et al.} 1991; see 
    below). The final images of SCAP 2000 had a resolution of about 1.5-2 
    arcmin and a limiting magnitude of about 13.5 for sources with a 
    spectrum similar to that of an A0 star. 
 
    SCAP-2000 surveyed about 15\% of the sky, and some results, pertaining 
    to galaxies and their derived star formation rates, were published by 
    Donas {\it et al.} (1987) and Buat {\it et al.} (1987, 1989). 
 
      \subsection{Wide-Field UV Camera } 
 
    This instrument flew in December 1983, on the same SPACELAB-1 flight 
    which carried FAUST (see below) on its first orbital flight. The 
    experiment was 1ES022 and was designed to image a 66$^{\circ}$ wide 
    field with a resolution of 5 arcmin with an all-reflective camera. 
    The WF-UVC was affected by the high 
    orbital background, just as was FAUST alongside it in the STS bay. 
    Despite this deficiency, due 
    mainly to the high straylight at the Shuttle altitude in that specific 
    orbit, the best WF-UVC 
    images reached m$_{193}$=9.3. A description of the WF-UVC and its 
    results was given by Court\'{e}s {\it et al.} (1984). An interesting 
    result was the detection of UV-bright stars along the bridge connecting 
    the Magellanic Clouds, an extension of the Shapley wing. 
 
        \subsection{ FOCA } 
 
    The FOCA experiment flies a 39 cm diameter telescope on a balloon 
    gondola to altitudes higher than 40 km. There are two optical assemblies, one 
    yielding a field of view of 1$^{\circ}$.5 and another with 
    2$^{\circ}$.3, with image resolution of 10"-20". The flights take place 
    from a French launching ground (Aire-sur-Adour) and end a few hours 
    later in Italy. 
 
    The detectors were, as for SCAP 2000, image converter-intensifiers, and 
    the image was recorded on film. The measurements were done by PDS-ing 
    the films, with proper calibration of the intensities. The problem with 
    both SCAP 2000 and FOCA were altitude changes of the balloon, which 
    modified the bandpass of observation because of the influence of the 
    atmospheric transparency. The FOCA experiment was described by Milliard 
    {\it et al.} (1991). The bandpass excludes contributions larger than 
    10\% from the near UV to the visible for objects hotter than the Sun, 
    provided the balloon altitude is higher than 3-5 mbar and the zenith 
    distance less than 55$^{\circ}$ (Laget {\it et al.} 1991b). 
 
    FOCA surveyed some 70 square degrees of the sky. Some results on 
    galactic metal-poor globular clusters were reported by Laget {\it et 
    al.} (1991a, 1991b). One interesting feature found by the FOCA imaging 
    is a small dark patch observed in visible against M13, which is 
    considered to be a foreground dark cloud, but which does not show up in 
    the UV (Laget {\it et al.} 1991a). FOCA results on UV images of nearby 
    galaxies were reported by Vuillemin {\it et al.} (1991), and analyzed 
    (among others) by Court\'{e}s {\it et al.} (1993, NGC 4258), Buat {\it et 
    al.} (1994, M33), Bersier {\it et al.} (1994, M51), Reichen {\it et al.} 
    (1994, M81) and Petit {\it et al.} (1996, M51). A study of UV galaxies 
    in the Coma cluster was published by Donas {\it et al.} (1995). Galaxy 
    counts and color distributions for objects in the magnitude range 15.0 
    to 18.5 were published by Milliard {\it et al.} (1992); these served as 
    basis for a prediction of UV galaxy counts by Armand \& Milliard (1994), 
    which requires more late-type galaxies than predicted by optical data, 
    or faster evolution of the galaxies. 
 
    The LAS group have a parallel ground-based 
    observational follow-up program, for systematic identification of their 
    UV sources. Some of their UV sources are fainter than the limit of 
    POSS-I, requiring POSS-II data and dedicated observations at large 
    telescopes, conducted (with collaborators) at Palomar, Keck and WIYN 
    telescopes. This confirms the identification as blue galaxies, and their 
    high projected density, which apparently reaches out to z=0.68 (Milliard 
    1996, private communication; Martin 1997). The 1$^{\circ}$.5 field centered on the 
    Abell 2111 cluster of galaxies contains $\sim$450 galaxies and $\sim$350 
    stars. The partial redshift survey mentioned above indicates that most 
    galaxies are in the fore- and background, only $\sim$20\% of them 
    belonging to the cluster. A similar study on the FOCA field centered on SA57
    identified 45 sources with m$_{UV}\leq$18.5 with 40 galaxies, 3 QSOs and
    two stars (Treyer \etal 1997). The galaxies are mostly later than
    Sb and their UV emission is interpreted as indicating enhanced star 
    formation rates in these objects.

    During some test flights of FOCA in 1996 and 1997 the telescope was equipped
    with a 25 mm diameter detector which has electronic readout (resistive anode). No
    results have been reported yet from these flights. FOCA shall be upgraded with a 
    40 mm diameter detector, which matches better its focal plane, and which shall
    have a crossed-delay-line anode readout for high angular resolution observations.
 
        \subsection{ FUVCAM } 
 
    The NRL group headed by G. Carruthers flew a number of far-UV wide-field 
    imagers (FUVCAMs) on rockets (Carruthers {\it et al.} 1978, 1980), which 
    observed 
    M31 and the North American Nebula. These flights used the Mark II 
    FUVCAM, an electrographic Schmidt camera with a field of view of 
    11$^{\circ}$ and a (theoretical) resolution of 30". 
 
    FUVCAMs flew on a number of rocket flights, and on a longer-duration 
    space flight from 28 April to 6 May 1991, when it operated from the bay of 
    the Space Shuttle Discovery (STS-39). 
    The experiment was constructed by the UV astronomy group of the Naval 
    Research Laboratory (G. Carruthers and collaborators). It consists of 
    two electrographic cameras mounted side-by-side and bore-sighted. Each 
    camera covers a field of view of 10$^{\circ}$.5 square, with a final 
    resolution of $\sim$3 arcmin.

    The images were recorded on film, 
    which was later digitized. In order to overcome the limited dynamic 
    range problem, each field was imaged three times, with exposure times 
    increasing by about 3 to 10 times. The camera comprised two electronographic
    devices, one for the 105-160 nm band (123-160 nm with a CaF$_2$ filter),
    and the second for the 123-200 nm band (165-200 nm with a SiO$_2$
    filter). The experiment was described by Carruthers 
    et al (1992) and its calibration is described by Carruthers {\it et al.} 
    (1994). 

    The results from the FUVCAM observations were published in a series of 
    papers dealing with individual fields: Monoceros (15 November 1982 
    rocket flight, Schmidt \& Carruthers 1993a, 1994), Orion (6 December 
    1975 and 15 November 1982 rocket flights, Schmidt \& Carruthers 1993b), 
    and Sagittarius and Scorpio (Shuttle flight, Schmidt and Carruthers 
    1995). Not all flights used exactly the same optical configuration, however the 
    field of view was always very wide. 

    Generally, the limiting magnitude of FUVCAM is m$_{UV}$=9-10 mag  
    and the observing band was always shorter than 200 nm and longer than 
    Ly$\alpha$. The identification of UV sources was done routinely by correlating 
    against the SIMBAD data base. Thus, typically about 60\% of the sources 
    were identified and between 24 and 40\% of all sources could be 
    attributed to blends of early-type stars. Among the identified sources, 
    about half are early-B stars. The raw, uncalibrated FUVCAM images from
    the STS-39 flight are available on the web.
 
       \subsection{ GSFC camera (UIT prototype)} 
 
    A wide-field UV imager was flown by the Goddard Space Flight Center on a 
    number of rocket flights and with different focal plane assemblies. The 
    31 cm diameter telescope had a field of view of 11$^{\circ}$.4 and its 
    spatial resolution was about 50" FWHM. The detector was, in all cases, a 
    UV sensitive cathode with an MCP intensifier coupled to film. 
 
    Bohlin {\it et al.} (1982) described the instrument and its observations 
    of the Orion nebula in four spectral bands (140, 182, 224, and 262 nm) 
    obtained during a flight on 11 December 1977. Observations from a flight 
    on 21 May 1979, when M51 was observed, are described by Bohlin {\it et 
    al.} (1990). 
 
        Smith \& Cornett (1982) reported observations of the Virgo cluster 
    with this telescope, where the response band peaked at 242 nm and was 
    about 110 nm wide. The sensitivity, for the Virgo cluster exposure 
    obtained during a flight on 22 May 1979, reached m(242 nm)=16.3. 
 
    The experiment also imaged the LMC (Smith {\it et al.} 1987) in two UV 
    bands, 149.5 and 190 nm with FWHM of 20 and 22 nm respectively. In the 
    LMC, Smith {\it et al.} measured UV fluxes from 122 stellar 
    associations, from which they derived a model for the progress of the 
    star formation process in this very nearby dwarf galaxy.

       \subsection{ GLAZAR } 
 
    A 40 cm telescope operated briefly on the Mir space station. This is the 
    GLAZAR-2 telescope, described by Tovmassian {\it et al.} (1991a), a 
    direct follower of the GLAZAR-1 telescope (Tovmassian {\it et al.} 
    1988). Its existence and results were described mainly in Soviet 
    publications, thus it was not well-known in the West. The limiting 
    magnitude of GLAZAR-1 at 164 nm was originally 11 mag, but it steadily 
    declined by about 2.5 mag from the launch and operation start in 1987 
    during the subsequent 2.5 years (Tovmassian {\it et al.} 1991b). This 
    was probably a detector-related problem (Tovmassian, private 
    communication). A few scientific results were reported in this 
    describing paper. 
 
    The two GLAZAR telescopes have identical optical designs. They are 40 cm 
    Ritchey-Chr\'{e}tien designs, imaging a 1$^{\circ}$.3 field of view onto 
    a 40 mm multi-channel plate intensifier with phosphor output. The 
    instantaneous image quality is $\sim$10" and the images are recorded on 
    film. The spectral range of observation is determined by a filter close 
    to the focal plane, with $\lambda_c\simeq$164 nm and 
    $\Delta\lambda\simeq$25 nm. The filter acts as a vacuum barrier between 
    the telescope tube (open to space) and the detector-film transport 
    mechanism. The film is exchanged with the telescope in a 
    parked position against a small airlock. 
 
    GLAZAR-1 was rigidly fixed to the MIR station; the low stability of this 
    platform yielded images of $\sim$40" FWHM, thus the low sensitivity of 
    the detections. GLAZAR-2 was mounted on double gimbals and was equipped 
    with coarse and fine star trackers, allowing access to wider sky areas 
    and better image quality through independent tracking of the telescope. 
    This allowed it to achieve $\sim$10" resolution. The 
    film was retrieved from the MIR by the supply spaceships, was developed, 
    digitized, and analyzed. Unfortunately, the film transport mechanism and 
    the airlock handle were damaged soon after launch and no results were 
    obtained. Since then ($\sim$1990), the GLAZAR-2 telescope has not been used. 
 
    Results from the GLAZAR flights, having to do with distribution and 
    identification of blue stars, were reported by Tovmassian {\it et al.} 
    (1991c, 1992, 1993a, 1993b, 1994a, 1994b, 1996a). The latter, in 
    particular, emphasizes the decrease in sensitivity of the GLAZAR; only 
    217 stars were detected in a field some 12 square degrees in size, with 
    a limiting monochromatic magnitude of m$_{164}<$8.7 mag. The observations 
    concentrated on OB associations. About 12 different celestial directions were 
    observed and in each case an area of 10-20 degrees$^2$ was imaged. On 
    all exposures a total of 489 stars were measured. 
 
       \subsection{FAUST} 
 
    FAUST is the Fus\'{e}e Astronomique pour l'Ultraviolet Spatiale, or the Far 
    Ultraviolet Space Telescope. It started life as a Wynne telescope 
    (reversed optics: the secondary being larger than the primary), whose 
    purpose was to image the UV sky. The instrument was described by 
    Deharveng {\it et al.} (1979) and was originally intended for a rocket 
    launch. This version of the instrument was used for at least two 
    successful flights, mainly used for calibrations. 
 
    The field of view of FAUST was $\sim8^{\circ}$ and the angular resolution was 
    1-2 arcmin. On SPACELAB-1, during a flight on board the Space Shuttle in 
    December 1983, FAUST did not succeed to obtain significant data, 
    although 47 exposures of 22 targets were obtained on film, behind an 
    image converter-intesifier. Unfortunately, the on-orbit background was 
    very high and almost no objects were recorded. From the few exposures 
    which could be analyzed it is worth mentioning an interesting image of 
    the Cygnus Loop, published by Bixler {\it et al.} (1984). 
 
    For its second orbital flight FAUST was equipped with a new detector, 
    which incorporated a CsI cathode, a three-stage chevron arrangement of 
    MCPs, and a novel wedge-and-strip anode. The flight took place on board 
    the Shuttle Atlantis in March 1992, and was described by Bowyer \etal 
    (1993). During this flight FAUST telemetered every detected photon. This 
    allowed the rejection of spurious photons, originating from the firing 
    of the Shuttle attitude jets or from high orbital background. FAUST 
    observed 22 fields, among which were the North Galactic Pole, the Virgo 
    cluster of galaxies, and other galaxy clusters and regions of interest. 
    FAUST's observations yielded a catalog of 4698 UV sources (Bowyer \etal 
    1995), measured in a 
    band about 30 nm wide which was centered at 165 nm. The band was defined 
    by the CsI cathode and a CaF$_2$ window with multi-layer coatings. The 
    detection was by an impartial automatic algorithm, and the 
    identification was through correlations with existing catalogs. 
 
    Selected results were published from the FAUST imagery obtained during 
    the second flight. Deharveng {\it et al.} (1994) analyzed the UV 
    emission of galaxies. Haikala {\it et al.} (1995) imaged a galactic 
    cirrus cloud and showed the good correlation between the UV diffuse 
    emission and the IRAS 100$\mu$m emission, from which they constrained 
    the albedo {\bf a} and the isotropy parameter {\bf g} of the dust 
    particles. Sasseen {\it et al.} (1995) used the spatial power spectrum 
    of FAUST images to search for an extragalactic UV background component. 
    Sasseen \& Deharveng (1996) correlated the UV background detected
    by FAUST with the 100$\mu$m FIR emission measured by COBE/DIRBE.
 
    An interesting result was published by Court\'{e}s {\it et al.} (1995). It 
    is a confirmation of the UV extension of the Shapley wing of the SMC, 
    first detected with the WF-UVCAM (Court\'{e}s {\it et al.} 1984). The FAUST 
    observations are deeper (m$_{UV}<$13.9) and have better angular 
    resolution than those of the WF-UVCAM. The observations confirm the 
    earlier findings that the Shapley wing contains a population of young 
    stars formed at most 5 Myrs ago with an IMF which is flatter than that 
    of the stars in the SMC core (M$_{up}<30$M$_{\odot}$), {\it i.e.,} later than 
    mid-O type. The FAUST observations of the SMC, combined with FIR, HI and 
    H$\alpha$ data, form part of the thesis of Okumura (1993).  
 
    A program to systematically investigate the FAUST data set takes place 
    at Tel Aviv University. We re-detect sources with a different automatic 
    algorithm, based on local S/N ratio. After attempting correlations with 
    existing catalogs, we apply astrophysical criteria to identify the 
    sources. About 10-15\% of the sources remain unidentified after this 
    procedure. These are identified with possible counterparts on the 
    Palomar Sky Survey and the counterparts are observed from the Wise 
    Observatory. To date, we analyzed completely five FAUST fields, the 
    North Galactic Pole (Brosch {\it et al.} 1996a), three fields 
    covering most of the Virgo cluster (Brosch {\it et al.} 1997), and
    one image in the direction of Coma (Brosch \etal 1998). Analysis 
    of other fields (Ophiuchus and other southern  fields) is very advanced. 
 
    The main results concern the distribution of UV stars, and in the case 
    of the Virgo cluster, also measurements of some 90 galaxies. In the Coma
    field we identified a large population of hot evolved stars, which are
    probably connected with the open cluster Mel 111. In total, 
    the FAUST frames analyzed so far at Tel Aviv yielded $\sim$100 
    galaxies and a few hundred stars.

        \subsection{UIT} 
 
    The Ultraviolet Imaging Telescope (UIT) was described by Stecher \etal 
    (1992). It consists of a 38 cm diameter Ritchey-Chr\'{e}tien telescope 
    with a 40 arcmin field of view (about 200 times larger than the WFPC2 of 
    the HST). The telescope was mounted on the Instrument Pointing System in 
    the Space Shuttle bay for the ASTRO-1 (December 1990) and ASTRO-2 
    flights (March 1995). In addition to the tracking capabilities of the 
    IPS used in the ASTRO flights, the UIT was equipped with an articulated 
    secondary mirror, which provided even finer tracking. 
 
    The focal plane consisted of two image converter-intensifiers, one for 
    the far-UV and the other for the mid-UV range. The image tubes were 
    coupled to film on which the images were registered. The film frames 
    were digitized to a 2048$\times$2048 pixel format, and to enhance the 
    dynamic range, multiple exposures of different duration were taken 
    through each filter. During the ASTRO-1 flight more than 800 exposures 
    were taken. The evaluation of the UIT results from the ASTRO-1 flight 
    indicates a resolution of about 3" and a sensitivity sufficient to 
    register stars of UV monochromatic magnitude 19.5 (for a hot, unreddened 
    source: O'Connell 1992). However, for an indication of the actual 
    sensitivity, see below. 
 
    First results from the UIT mission were published in a 1992 dedicated 
    volume of the Astrophys. J. Letters (vol. 395). The papers discuss the 
    UV scattering properties of dust in NGC 7023 (Witt {\it et al.} 1992), 
    observations of the Cygnus Loop (Cornett {\it et al.} 1992), of the Crab 
    Nebula (Hennessy {\it et al.} 1992), of globular clusters such as M79 
    (Hill {\it et al.} 1992a) and $\omega$ Cen, M3 and M13 (Landsman {\it et 
    al.} 1992), of SN 1987A (Crots {\it et al.} 1992), the SN environment 
    and 30 Doradus (Cheng {\it et al.} 1992), of the association NGC 206 in 
    M31 (Hill {\it et al.} 1992b), of M81 (Hill {\it et al.} 1992c), of NGC 
    628 (Chen {\it et al.} 1992), of M33 (Landsman {\it et al.} 1992), of 
    nearby galaxies (O'Connell {\it et al.} 1992), and of NGC 1275 (Smith 
    {\it et al.} 1992). 
 
    Most of the final results from both UIT flights have not yet been 
    published, but some were reported at meetings (notably, at AAS 
    meetings). I mention in particular published studies of nearby galaxies 
    (M31: Hill {\it et al.} 1995a, and Magellanic Clouds: Hill {\it et al.} 
    1993, 1994, 1995b), mostly from the ASTRO-1 flight. Pica {\it et al.} 
    (1993) mention a catalog of $\sim$2,200 sources observed near 250 nm 
    derived from images of 66 fields obtained in the ASTRO-1 flight. Of 
    these, about 300 did not have counterparts in published catalogues. The 
    catalog (Smith {\it et al.} 1996) covers 16 square degrees of the sky 
    and contains 2,244 objects culled from 48 pointings. The identification 
    was done through correlations with optical catalogs and, in fact, most 
    sources are from the HST Guide Star Catalog. The percentage of 
    identified sources is 88\%, mainly because of the good spatial 
    resolution of the images. 
 
    Unfortunately, the UIT catalog was based only on the near-UV (165-290 
    nm) observations of the ASTRO-1 mission; during the ASTRO-2 mission this 
    camera failed and no observations were obtained. The catalog lists the 
    sources obtained in the 48 selected fields, where the typical exposure 
    depth (though by no means the completion limit) was 17.2 mag. 
 
    From the ASTRO-2 flight, Neff {\it et al.} (1995) reported that some 30 
    peculiar galaxies have been observed in one or two UV bands (B1: 125-180 
    nm and/or B5: 140-180 nm). The galaxies include interacting, starburst 
    and/or active objects. Smith {\it et al.} (1995) reported at the same 
    meeting that five clusters of galaxies were observed during the ASTRO-1 
    flight, and additional clusters were included among the targets observed 
    during the ASTRO-2 flight. A recent interesting paper (Hill {\it et 
    al.} 1997) combines ASTRO-2 UIT observations at 152 nm and optical 
    imagery to determine the colors and extinctions of HII regions in M51. 
    The authors find that the total-to-selective extinction A(152)/E(B-V) in 
    M51 increases with radius (or with decreasing metallicity). In addition, they
    find that the H$\alpha$ flux is depleted in the inner regions of the
    galaxy; this they interpret as increased Lyman continuum extinction. Parker \etal 
    (1996) studied a few OB associations in LMC using ASTRO-2 data at 152 
    nm. 
 
    O'Connell \& Marcum (1996) remarked, from UIT images of galaxies, that a 
    comparison of visible and UV images of the same nearby galaxies 
    indicates a trend from normal to abnormal, galaxies turning into later 
    morphological types with a higher incidence of irregular galaxies as the 
    wavelength of observation gets bluer. A similar claim was also made by 
    Giavalisco {\it et al.} (1996), based on simulated HST images using UIT 
    images of nearby galaxies. Other papers on galaxies, resulting from UIT 
    imagery are by Waller \etal on M101 (1997), by Smith \etal on NGC 3310 
    (1996b), on the SMC by Cornett \etal (1997), and a number of papers in 
    the proceedings of the Seventh Astrophysics Conference (Holt \& Mundy 
    1996) by Waller \etal, Fanelli \etal, Neff \etal, Hill \etal, Smith 
    \etal and O'Connell.  
 
    The UV images of nearby galaxies obtained by UIT are the baseline 
    templates showing how galaxies appear in the UV, with which one can 
    begin to understand the images of distant galaxies taken by HST in 
    optical bands; these correspond to rest-frame UV, proving the comparison 
    valid (Giavalisco \etal 1996). 
 
\subsection{UVISI}

     The UVISI instrument operated on-board the Mid-Course Space 
    Experiment (MSX) satellite. The mission was primarily 
    military in character (BMDO), aiming at detecting and 
    characterizing sources of UV emission (or atmospheric opacity), which 
    could affect the detection, identification, and tracking of missiles and 
    warheads. As side-benefits, these missions will produce full or partial 
    sky surveys in the UV. MSX was launched on 24 April 1996 and the 
    operation of UVISI started in 1997 and ended in early 1998. 
 
    The UVISI instrument consists of five spectrographic imagers and
    four imagers (Carbary \etal 1994). The narrow-field UV imager is of 
    interest to UV sky surveys. It images a 1$^{\circ}.6\times1^{\circ}.3$ 
    field of view with a resolution of $\sim20$" in one of three spectral bands
    from 180 nm to 300 nm (180-300, 200-230, or 230-300 nm; J. Murthy, private
    communication). There is also a wide field imager, with a FOV of  
    10$^{\circ}\times13^{\circ}$ and a resolution of $\sim$3'. The detectors are 
    CCDs coupled through fiber-optic tapers 
    to the phosphor outputs of image intensifiers. Based on the description 
    of UVISI (Heffernan {\it et al.} 1996), the narrow field UV imager with 
    no filters is sensitive to sources which produce 2 photons/cm$^2$/sec; 
    this should be equivalent to a limiting magnitude of 13.9 (monochromatic, 
    at 
    the band center, 240 nm). However, J. Murthy communicated an
    effective sensitivity limit of m$_{UV}$=20.0.
    The results of the UVISI observations
    have not yet been put in the public domain.

        \subsection{HST } 
 
    Although far from being a survey instrument, the HST has some UV 
    capability with its WFPC-1 or WFPC-2 cameras, and with the FOC 
    instrument. However, the long-wavelength rejection of light by the WFPC 
    filters is not fully satisfactory, with the notable exception of the
    F160BW ``Woods'' filter. Far UV observations suffer from red 
    leaks, requiring complicated compensatory measurements. The FOC has 
    better rejection of optical light because of its cathode. Both cameras 
    have small fields of view: 2'.6 for WFPC-1, 2'.5 for WFPC-2 (total
    sky coverage 5.02 arcmin$^2$), 44" for the 
    pre-COSTAR FOC, and 28" for the post-COSTAR FOC (7" with full sampling 
    of the PSF and with $\sim$1.55\% efficiency). An example of the UV 
    capability of the FOC is the snapshot survey of nuclei of galaxies (Maoz 
    {\it et al.} 1996), where circumnuclear star-forming rings were 
    identified in a few objects through 230 nm imaging. 
 
    The new instrument STIS has significant UV capabilities with the MAMA 
    detectors. These have a throughput higher by one order of magnitude or 
    more than the WFPC-2 with UV filters (F160BW or F170W), but their 
    largest field of view is only 25"$\times$25". In particular, the 
    far-UV (FUV) MAMA with the Csi cathode and SrF$_2$ short band cutoff 
    filter allows
    one to reach very low background values while retaining reasonable 
    throughput ($\sim$2.5\% at peak). STIS with the FUV MAMA and the SrF$_2$
    cutoff filter have a throughput higher by $\sim27\times$ than that
    of WFPC-2 with the F160BW filter. This ``almost'' compensates
    for the field-of-view, which is smaller by $\sim29\times$ than
    that of the WFPC-2, disregarding the region vignetted when
    the F160BW filter is used.   
 
    The next servicing mission for the HST, currently scheduled for December
    2 1999, will see the installation of the 
    Advanced Camera for Surveys (ACS). This instrument is designed with 
    three separate channels, of which one has a UV capability and another is 
    a solar-blind channel. The High Resolution Camera shall cover a field of 
    view of 27"$\times$26" with a plate scale of 0".025/pixel using a 
    1024$\times$1024 CCD. The spectral range covered shall be from 200 to 
    1000 nm with a net efficiency of $\sim$15\% in the UV part of the band. 
    The Solar Blind Camera shall cover a FOV of 33"$\times$30" 
    (1.6$\times$ that of the STIS FUV MAMA) with a plate 
    scale of 0".03/pixel using a spare MAMA detector from STIS, with a net 
    efficiency of $\sim$4\% at 140 nm. 

    The latest instrument (for the time being) selected for an HST upgrade
    is the Cosmic Origins Spectrograph (COS). This shall replace COSTAR
    during the fourth HST servicing mission in 2002. It is an instrument
    optimized for high-throughput spectroscopy of point sources in the
    115-205 nm band.  The high throughput is achieved by minimizing the
    number of reflections between the aperture and the detector (a single
    reflection, at the grating), allowing more than one order of magnitude
    improvement in this aspect relative to STIS. The spectral resolution
    can be high (R$\simeq$20,000-24,000) or intermediate 
    (R$\simeq$2,500-3,500). The esimated sensitivity is such that, in high
    resolution mode, a source with monochromatic magnitude 15.6 will
    produce a spectrum with S/N=10 per resolution element in 10,000 sec.

    Based on the ``survey 
    power'' parameter $\theta$ to be described below, UV imaging by the HST 
    has significant survey potential, despite the small sky area it can 
    cover. In fact, while a STIS-based survey would only be $\sim1.6\times$ more 
    effective than one with the WFPC-2, a survey with the ACS/SBC would be
    $\sim45\times$ better.

       \subsection{Background measurements} 
 
    Attempts to measure the diffuse UV background in the TD-1 era are lumped 
    together in one section. They consist of observations by the UV 
    photometer on the Apollo-Soyuz mission (Paresce {\it et al.} 1980), the 
    measurements of the D2B-Aura satellite (Maucherat-Joubert {\it et al.} 
    1980, Joubert {\it et al.} 1983, Lequeux 1982), sky measurements from a 
    rocket flight (Jakobsen {\it et al.} 1984), and from the Dynamics 
    Explorer satellite (Fix {\it et al.} 1989), and by the two Shuttle-borne 
    UVX instruments from JHU and Berkeley (Murthy {\it et al.} 1989; Hurwitz 
    {\it et al.} 1989). In addition, observations done for other purposes 
    but used to derive the UV background were by Zvereva {\it et al.} 
    (1982), Weller (1983), Onaka (1990), and Jakobsen {\it et al.} (1984). 
    The UVX observations yielded one of the lower background values at
    160 nm: 280$\pm$35 c.u. (Martin \etal 1991).

    The FAUST instrument has been used to derive the UV background, with
    the emphasis on an attempt to disentangle the Galactic from the extragalactic 
    signal based on spatial power spectra (Sasseen \etal 1995). Their best
    explanation for the signal detected is that it is due to starlight
    scattered off dust grains. This is apparently confirmed by the correlation
    between the UV background measured by FAUST and the 100$\mu$m FIR
    emission measured by COBE/DIRBE (Sasseen \& Deharveng 1996).
 
    The observations of UIT have also been used to analyze the UV background. 
    Waller {\it et al.} (1995) quote orbital night-time background levels of 
    $\sim$3000 ph/sec/cm$^2$/ster in the near-UV band ($\sim$250 nm) and 
    $\sim$5000 ph/sec/cm$^2$/ster in the far-UV band ($\sim$150 nm). After 
    correcting for instrumental and orbit-dependent effects, and 
    compensating for galactic diffuse UV radiation through a correlation of 
    UV and FIR diffuse emission, Waller {\it et al.} identified a possible 
    extragalactic component of 200$\pm$100 c.u. As will be discussed later, 
    the UIT background measurements and some deep Voyager UV spectra are now the 
    strongest constraints for a cosmologically interesting UV background. It 
    appears that these constraints conflict extrapolated UV observations of 
    faint galaxies. 
 
        \section{ Modern EUV observations } 
 
        \subsection{ ROSAT WFC } 
 
    The EUV sky was explored almost simultaneously by two instruments. The 
    ROSAT EUV Wide Field Camera was an add-on instrument to the ROSAT X-ray 
    all-sky survey satellite. Its first results, and a description of the 
    instrument, were published by Pounds and Wells (1991). The camera consists
    of a nested three-mirror Woltjer-1 telescope with a MCP detector equipped
    with a CsI cathode at the common focus of the mirrors. Selectable
    filters allow observations in four spectral bands: 52-73 nm (17-24 eV), 
    14.9-22.1 nm (56-83 eV), 11.2-20 nm (62-111 eV), and 6-14 nm (90-210 eV). 

    The first EUV 
    all-sky survey was performed by this instrument during 1990-1991 (Pye 
    1995). The survey was done in two bands: S1 (6-14 nm) and S2 (11.2-20 
    nm). The effective spatial resolution was 3 arcmin and the source 
    location was good to within one arcmin. The initial results were 
    reported by Pounds {\it et al.} (1993) as the WFC Bright Source Catalog 
    (BSC). The reprocessed data, with more sources, were published by Pye 
    {\it et al.} (1995) as the 2RE catalog. 
 
    The 2RE catalog contains 479 EUV sources, 120 more than the BSC, 
    observed with a median exposure of about 1600 sec. Of the entire 2RC 
    catalog 52\% of the sources are identified as active, late-type (F, G, 
    K, and M) stars; 29\% are hot white dwarfs, and less than 2\% are 
    extragalactic sources (AGNs). Thirty-four sources (about 7\%) were still 
    unidentified in 1995 (Pye {\it et al.} 1995). The cumulative source 
    distribution indicates that the late-type stars dominate the source 
    counts at faint count rates ($<$0.02 cps), surpassing the contribution 
    from white dwarfs. 
 
        \subsection{ALEXIS } 
 
    The region bordering the EUV and the X-rays was explored by the ALEXIS 
    spacecraft (Priedhorsky 1991). ALEXIS stands for ``Array of Low Energy 
    X-ray Imaging Sensors'' and consists of an array of six wide-field small 
    telescopes with multi-layer coated mirrors with prime-focus imaging detectors, 
    and sensitive to radiation at 66, 71, or 93 eV (18.8, 17.2, and 13.3 nm). 
    The passbands are defined by the mirror coatings and by filters
    positioned in front of the detectors. Recent descriptions of 
    ALEXIS are by Bloch (1995) and  by Roussel-Dupr\'{e} \& Bloch (1996). 
 
    Unfortunately, during the Pegasus launch on 25 April 1993 the satellite 
    was damaged, lost one of its four solar panels, damaged the on-board
    magnetometer used for position-sensing, and entered an 
    uncontrolled spin. This reduced the available electrical power and 
    questioned the effectiveness of the survey. The spacecraft was 
    eventually stabilized and effective observations could be made from 
    mid-July 1993. ALEXIS points its six telescopes perpendicular to the 
    Sun-Earth line and scans the sky with them every 50 seconds. The 
    attitude of the satellite is known to better than 0$^{\circ}$.5 and the 
    information collected is received at a Los Alamos ground station. 
 
    The EUV telescopes of ALEXIS are arranged in three co-aligned pairs. 
    Each has a 33$^{\circ}$ field of view and a resolution of 0$^{\circ}$.25 
    (limited by the spherical aberration of the mirrors). The first sky maps 
    of ALEXIS were produced on 4 November 1994. Since 1995, the daily maps 
    are searched for EUV transient sources, with a 12-24 hour response time. 
    The transients are then serched for in the error ellipse of ALEXIS by 
    ground-based telescopes, and are followed up. Among the more interesting 
    EUV transients the more notable are the super-outburst of VW Hyi in June 1994, the 
    ALEXIS J1114+43=1ES 1113+432 outburst in November-December 1994, and the 
    fast transient ALEXIS J1139-685. In addition, outbursts 
    from U Gem, AM Her, and AR UMa were also detected. Some results on
    transient EUV sources were reported by Roussel-Dupr\'{e} \etal (1996).
 
    Using the pre-flight information, the ALEXIS team calculated that 
    $\sim$10\% of the brightest EUVE sources (see below) should be 
    detectable (Jeff Bloch, private communication), and apparently this is 
    the situation. Most of the sources are WDs, and it is expected that the 
    catalog from the first three years of operation will contain $\leq$50 
    sources.

        \subsection{EUVE } 
 
    The EUV sky was thoroughly investigated by the Extreme Ultraviolet 
    Explorer (EUVE) spacecraft, launched on June 7, 1992. (Bowyer \& 
    Malina 1991). The instrument 
    consists of three ``scanning'' telescopes, co-aligned and perpendicular to 
    the spacecraft spin axis, which have a 5$^{\circ}$ field-of-view, and a 
    deep survey spectrometer telescope looking along the spin axis and 
    pointing away from the Sun. The angular resolution is $\sim$1 arcmin. 
 
    EUVE mapped the sky in four spectral bands, from 7 to 70 nm (18 to 170 
    eV). The initial results were published as ``The First EUVE Source Catalog'' 
    (Bowyer {\it et al.} 1994), which contains 410 sources. A Second EUVE 
    Source Catalog has also been published (Bowyer {\it et al.} 1996). 
    For the foreseeable future, the EUVE catalogs set the standard in the 
    knowledge of the EUVE sky. 
 
    The second EUVE catalog includes additional observations to those from 
    the all-sky survey. In total, there are 734 sources, of which about 65\% 
    have optical, UV, X-ray, and/or radio counterparts. It is noteworthy 
    that 211 of these sources were never before observed in the EUV range. 
    The majority of the identified sources (55\%) appears to be late-type 
    stars (G, K, and M), which presumably have active chromospheres. 
 
    The observation that large numbers of late-type stars are detected as 
    EUV sources is 
    supported by the findings of the ``right angle'' survey, which has the 
    deepest sensitivity. These observations are typically 10$^4$-10$^5$ sec, 
    whereas the typical observations used for the derivation of the catalog 
    are $\sim$500 sec, thus the right angle survey can detect much weaker 
    sources than in the general survey. The late-type stars make up almost 
    half the sources in the right angle survey, whereas they are only less 
    than a third of the sources in the general sky survey. The late-type 
    stars are also the absolute majority among the EUV sources detected in 
    the ``ecliptic deep survey'', another sky region with very high exposure 
    ($\sim$20,000 sec/pixel over a 2$^{\circ}\times$180$^{\circ}$ area along the 
    ecliptic). Among all the sources, $\sim$8\% are identified with 
    extragalactic objects. Bowyer {\it et al.} (1996) compared the second 
    EUVE catalog with other EUV surveys. They find that 263 (of the 479) 2RE 
    sources appear in the new catalog, and note that the undetected sources 
    are either variable or belong to the ``unidentified'' category of faint 
    sources that may be spurious. 
 
    A new catalog, reaching sources fainter by $\sim$60\% than the second EUVE 
    catalog, has been produced by Lampton {\it et al.} (1997). The catalog 
    is based on coincident sources between the EUVE 10 nm list and the ROSAT 
    all-sky survey sources detected in the broadband event window (0.1-2 
    keV). It contains 534 EUV sources, of which 166 were not previously 
    discovered. Of these, 105 have been identified and 77\% of them are 
    late-type stars. White dwarfs and early-type stars make up only 
    $\sim$14\% of the sources, and there are no extragalactic objects at 
    all among them. 

    One important outcome from the EUVE mission is the first Spectral
    Atlas in the EUV range (Craig \etal 1997). The Atlas contains EUV
    spectra of 95 stars ranging from bright B stars to M dwarfs,
    white dwarfs, and cataclysmic binaries.
 
    \subsection{UVS, HUT and ORFEUS}  
 
    One should note that {\bf some} information about the EUV properties of 
    stars and galaxies was obtained with the Voyager Ultraviolet 
    Spectrometers (UVS), and with the Hopkins Ultraviolet Spectrometer (HUT) 
    and ORFEUS (Orbiting and Retrievable Far and Extreme Ultraviolet 
    Spectrometer) flights. While the UVS instruments use non-imaging optics 
    to record the EUV spectra of bright objects, HUT and ORFEUS used imaging
    telescopes to study spectra in the FUV range.  

    The generic UVS instrument installed on both Voyager spacecraft was 
    described by Broadfoot \etal (1977). It consists
    of a mechanically collimated objective grating spectrometer covering
    the range 50-170 nm with 1 nm resolution and was included in the
    Voyager missions primarily to study the composition and structure of
    the atmospheres of giant planets and their satellites. Due to the long
    cruise periods between planetary encounters, the Voyager UVS instruments 
    were uniquely able to make deep FUV observations of stars ({\it e.g.,}
    Chavez \etal 1995), galaxies (Alloin \etal 1995), 
    and the interstellar medium (Lallement 1993).

    HUT is a 0.9-meter telescope which operated from the Space Shuttle bay while
    mounted on the instrument pointer of ASTRO, alongside the UIT and WUPPE
    telescopes. A description of the instrument and its calibration was
    published by Davidsen \etal (1992). Briefly, HUT performed spectroscopy 
    in the 91.2-185 nm band with a resolution of $\sim$0.3 nm, and in the
    second order, in the 41.5-91.2 nm band with $\sim$0.15 nm resolution.
    The effective area of HUT ranged up to 10 cm$^2$ for the ASTRO-1 flight,
    and up to 25 cm$^2$ for the second flight. Note that HUT suffered a steady
    decrease of sensitivity at short wavelengths throughout the flight; a 25\%
    reduction at 80 nm from flight start to close to the end of the ASTRO-2
    flight. The results from the two flights of HUT were reported in some 60 
    papers, ranging from the 
    Moon's UV emission (Henry \etal 1995) to a search for the signature of 
    decaying heavy neutrinos (Davidsen \etal 1991).
 
    ORFEUS flew twice on the ASTRO-SPAS platform with a 1.0-meter telescope 
    and spectrographs covering the range 40-115 nm (R=3000) and 90-125 nm 
    (R=10,000). The same ASTRO-SPAS platform carried the Interstellar Medium 
    Absorption Profile Spectrograph (IMAPS; Princeton University). In this
    instrument, a 
    mechanical collimator restricts the FOV to $\sim$1$^{\circ}$ and feeds 
    an echelle spectrograph with R$\leq$120,000 in the range 95 to 115 nm. 
    This is intended for the study of ISM absorption profiles in the spectra 
    of bright stars.  Results from the IMAPS experiment were reported
    by Jenkins \etal (1996).
 
    OREFUS had $\sim$5 cm$^2$ effective area in the 40-90 nm band and 
    $\sim$9 cm$^2$ in the 90-120 nm. Its main results, from the first
    flight, include the discovery 
    of the S  and P elements in the photospheres of two white dwarfs 
    (Vennes \etal 1996), various studies of coronal gas ({\it e.g.,} 
    Hurwitz \etal 1995; Hurwitz \& Bowyer 1996) and an analysis of the 
    DO white dwarf HD149449B (Napiwotzki \etal 1995). An interesting 
    result was also the analysis of the O3If star HD93129A (Taresch
    \etal 1997); this object has a ZAMS mass of 120 M$_{\odot}$. The
    ORFEUS results from the second flight are being prepared for publication 
    in a special Astrophys. J. Letters issue.

        \section{Modeling the UV sky } 
 
    In parallel with observations by rocket and satellite instruments, 
    attempts to simulate the UV sky were made, {\it e.g.,} by Henry (1977). He 
    used a transformation from optical to UV based on Apollo 17 measurements 
    of bright, early-type stars and data from other experiments for cooler 
    stars at 148.2 nm. Data for other spectral regions were based on model 
    atmospheres from Kurucz {\it et al.} (1974). The transformations used 
    are described in Henry {\it et al.} (1975) and are essentially linear 
    relationships between the photon flux and the (B-V) color of a star for 
    all UV bands longward of Ly$\alpha$. Henry (1977) calculated that to 
    approximate the total UV starlight it is not necessary to observe very 
    faint stars; most of the UV light originates in relatively bright 
    (apparent) stars because of the influence of the interstellar 
    extinction. 
 
    Gondhalekar \& Wilson (1975) used a simple model for the distribution of stars 
    (plane-parallel distributions of different scale heights for various 
    types of O, B and A stars) to calculate the interstellar radiation field 
    between 91.2 nm and 274 nm. Gondhalekar (1990) used this model to calculate 
    the integrated UV emission by starlight in the Galaxy.
    The UV observations of TD-1 were used along 
    with model atmospheres to derive stellar properties in the UV. These 
    were combined with parameters of interstellar dust and the spectrum of 
    the diffuse UV background was derived. This was then compared 
    succesfully with measurements by Kurt \& Sunyaev (1968), Hayakawa {\it 
    et al.} (1969) and Henry (1972). 
 
    Two UV sky models were published in the last decade, by Brosch (1991) 
    and by Cohen (1994), later adopted for the FAUST bandpass by Cohen {\it 
    et al.} (1995). The Cohen (1994) model consists of a complicated model 
    of the galaxy, with a disk, a halo, the bulge, spiral arms and two local 
    spurs, as well as the Gould belt. The model of Brosch (1991) is a much 
    simpler adaptation of the Bahcall \& Soneira (1980) disk-bulge galaxy 
    model, to which the Gould belt was added, along with a thick disk of 
    white dwarfs, where the distribution followed the scale height of Boyle
    (1989). 
 
    The two models use different approaches in modeling the UV sky. While 
    Cohen (1994) creates absolute UV magnitudes for every class of objects 
    used by the model, Brosch (1991) calculates a transformation from 
    visible to UV using observed properties of IUE standard stars. The approach 
    used by Cohen employs model atmospheres from Kurucz (1991) and 
    integration across the filter bandpasses to yield the required [UV-V] color 
    indices. Later unpublished developments of the Brosch (1991) transformations
    use Kurucz (1992) stellar atmosphere models to extend the optical-to-UV
    transformation to later stellar types.
 
    Cohen (1994) tested successfully his model against the TD-1 results and 
    against those of the S201 Apollo 16 Moon telescope measurements. In a 
    later paper (Cohen {\it et al.} 1995) the predictions of the model were 
    tested against the detailed spectral distribution of stars in the North 
    Galactic Pole region identified by Brosch {\it et al.} (1996) in the 
    FAUST observation. 
 
    Similar encouraging results were obtained by Brosch (1991) in 
    comparisons with the S201 measurements. The visible-to-UV 
    transformation, derived by Brosch (1991) from IUE spectra, was later 
    used by Bilenko (1995) to verify a method of determining the 
    three-dimensional distribution of UV extinguishing ISM clouds. The 
    method uses the transformation to predict the apparent UV magnitude of a 
    star, based on its known optical properties as listed in the Hipparcos 
    Input Catalog (HIC). The expected UV magnitude is compared with that 
    measured by TD-1 and the total UV extinction is determined. Knowledge of 
    the distance modulus (from the HIC data) for an entire stellar 
    population in a given sky region, allows one to locate the extinguishing 
    dust clouds in 3D space. 

    In Figure 4
    I show how well can one predict the UV magnitude of a star,
    when the only known parameters are the V and B magnitudes. The comparison is
    done for stars in the direction of Virgo, appearing on three images obtained by the FAUST
    experiment. The optical information of magnitude and color is obtained from 
    the Tycho data set.

\begin{figure}[tbh]
\vspace{10cm}
\includegraphics{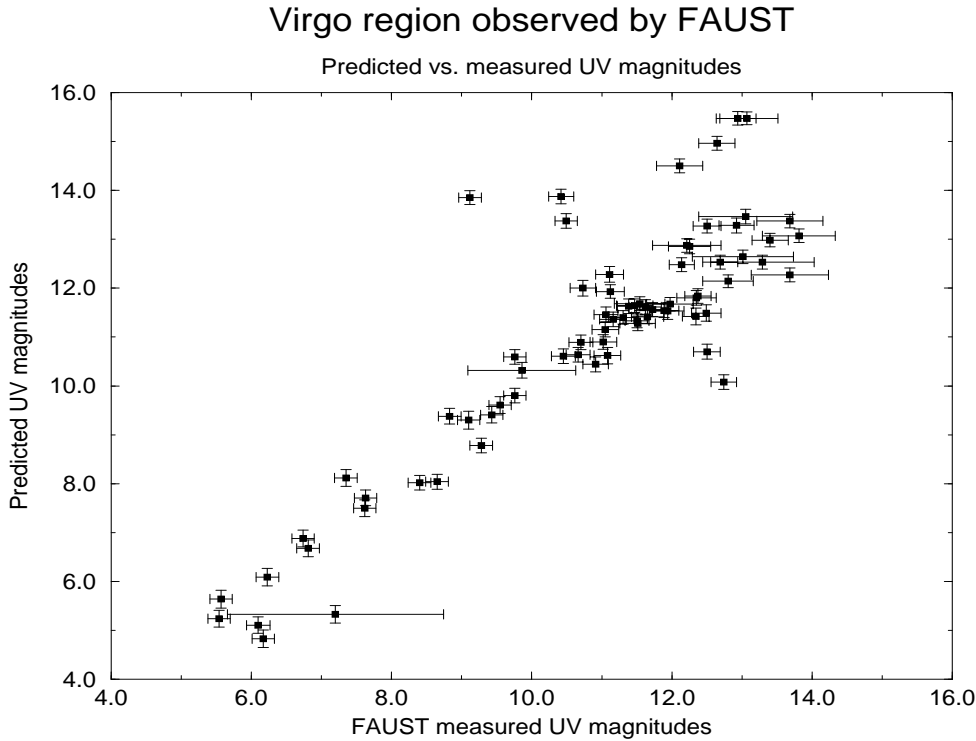}
\caption{Predicted {\it vs.} observed UV magnitudes at 165 nm for stars in the three 
FAUST fields on Virgo. The transformation relies on IUE observed spectra of stars
as well as on Kurucz (1992) model atmospheres.}
\end{figure}
 
    The diffuse UV sky background has been modeled by Murthy \& Henry 
    (1995). In this paper there are references to previous models accounting 
    for the UV background. Note though that not all published simulations 
    attempt to explain the faint UV sources. Extragalactic objects, such as 
    large or dwarf galaxies, AGNs, and QSOs, are not included in any 
    simulation. Nor has there ever been a satisfactory accounting for the 
    patchiness of the interstellar dust in the models, with the possible
    exception of a recent account by Witt \etal (1997), when analyzing the
    UV observations of FAUST.

        \section{Comparison of Survey Missions } 
 
    The various missions surveying the UV sky can be compared in terms of a 
    ``power'' parameter $\theta$ introduced by Terebizh (1986) and used by 
    Lipovetsky (1992) in a comparison of optical surveys:  
    \begin{equation}  
    \theta=\frac{\Omega}{4 \pi} 10^{0.6*(m_L-10)}  
    \end{equation}  
    where $\Omega$ is the sky area covered by the survey (4$\pi$ for TD-1) 
    and m$_L$ is the limiting (monochromatic) magnitude of the survey. 
    An all-sky survey to 
    a limiting magnitude m$_L\approx$8.5 (such as done by TD-1), has the 
    same ``survey power'' as a single HST WFPC-2 image exposed to show 21st mag 
    UV objects. The different UV and EUV missions discussed here are 
    compared in Table 3 in terms of their survey power. 
 
    A parameter similar to $\theta$, to compare missions whose goal was to 
    estimate the diffuse UV background, was introduced by Henry (1982): 
    \begin{equation} S=\frac{100}{[A \, \Omega \, \Delta\lambda]} 
    \end{equation} where A is the collecting aperture of the experiment, 
    $\Omega$ is the solid angle of its field of view, and $\Delta\lambda$ is 
    the spectral bandpass of the observation. Henry compared a number of 
    experiments and concluded that the Voyager UVS had the largest 
    $S$-parameter.   

    In order to evaluate the UIT performance as a survey instrument for 
    point sources, I assume that the ASTRO-1 flight imaged 66 fields and 
    ASTRO-2 flight another 100 fields, and that the depth of the UIT survey 
    was m$_{L}$=20. For the diffuse background case, I take the spectral 
    bandwidth to be 100 nm. Among the ``future missions'' categories I assume 
    that GIMI will ultimately conduct a full sky survey.  In 
    comparing TAUVEX and the UV/Optical monitor of XMM, I assume the same 
    mission duration and mode of operation for SRG and XMM. For HST, I 
    assume that 1000 independent WFPC-2 pointings with the F160BW filter, 1000
    with STIS FUV MAMA and SrF$_2$ and 1000 pointings with the ACS/SBC 
    (see below) could be assigned under a parallel imaging program. The
    sensitivity of the ACS/SBC is assumed to be slightly higher than
    that of the  STIS FUV MAMA, because of ther educed number of reflections.

    \begin{table}[htb]  
    \begin{center}  
    \caption{ UV and EUV survey missions}  
    \begin{tabular}{cccrrrrl}  
    \hline  
    Mission & Year & $\Omega$ (ster)& m$_L$ & $\theta$ & $\lambda\lambda$ 
    (nm) & N$_{sources}$ & Notes \\  
    \hline  
    TD-1 & 1968-73 & 4$\pi$ & 8.8 & 0.19 & 150-280 & 31,215 & 1 \\  
    S201 & 1972 & 0.96 & 11 & 0.30 & 125-160 & 6,266 & \\  
    WF-UVCAM & 1983 & 1.02 & 9.3 & 0.03 & 193 & ? & \\  
    SCAP-2000 & 1985 & 1.88 & 13.5 & 18.9 & 200 & 241 & 2 \\  
    GUV & 1987 & 5 10$^{-3}$ & 14.5 & 0.2 & 156 & 52 & Pointed phase \\  
    GSFC CAM & 1987+ & 0.03 & 16.3 & 14.4 & 242 & $\sim$200 & Virgo 
    observation \\  
    FOCA & 1990+ & 0.02 & 19 & 377 & 200 & $\sim$4,000 & Estimated \\  
    UIT-1 & 1990 & 3.8 10$^{-4}$ & 17 & 0.48 & $\sim$270 & 2,244 & UIT 
    catalog \\  
    GLAZAR & 1990 & 4.4 10$^{-3}$ & 8.7 & 6 10$^{-4}$ & 164 & 489 & \\  
    FUVCAM & 1991 & 0.09 & 10 & 7.5 10$^{-3}$ & 133, 178 & 1,252 & 3 \\  
    FAUST & 1992 & 0.33 & 13.5 & 3.3 & 165 & 4,698 & \\  
    UIT 1+2 & 1990, 95 & 1.3 10$^{-3}$ & 19 & 26 & 152-270 & 6,000 ? & 4 \\  
    \hline HST WFPC & 1990+ & 4.3 10$^{-4}$ & 21.0 & 134.8 & 120-300 & 50,000 ? 
    & 5 \\  
    HST STIS & 1997+ & 1.5 10$^{-5}$ & 23.8 & 222.7 & 120-200 & 10,000 ?\\   
    HST ACS/SBC & 1999+ & 2.4 10$^{-5}$ & 24.0 & 480 & 120-170 & 10,000 ?\\   
    \hline  
    MSX UVISI & 1997-98 & $\sim$0.01 & 18.0 & 50 & 180-300 & ? & 6 \\  
    ARGOS GIMI & 1998+ & 4$\pi$ & 13.6 & 136 & 155 & 2.5 10$^5$ & 7 \\  
    SRG TAUVEX & 1999+ & 0.06 & 19 & 1200 & 135-270 & 10$^6$ & 8 \\  
    XMM OUVM & 1999+ & $\sim$6 10$^{-3}$ & 19 & $\sim$100 & 185-600 & 10$^5$ ? & \\ 
    GALEX & 2002+ & 4$\pi$ & 19.4 & 4.4 10$^6$ & 130-300 & 2 10$^7$ & 9  \\ 
    \hline WFC & 1992  & 4$\pi$ & - & - & 10, 16 & 479 & \\  
    ALEXIS & 1994+ & 4$\pi$ & - & - & 13-19 & 50 ? & \\  
    EUVE & 1992  & 4$\pi$ & - & - & 7-70 & 734 & 10 \\  
    \hline  

    \end{tabular}  
    \end{center}

        Notes to Table 3: \\ 
        1: The unpublished extended version has 58,012 sources. \\ 
        2: 92 stars (Laget 1980) and 149 galaxies (Donas {\it et al.} 1987). 
    \\ 
        3: Only the Sag and Sco fields (Shuttle flights) included. \\ 
        4: Assumes 66 pointings for ASTRO 1 and 100  for ASTRO 2.\\ 
        5: Assumes 1000 observations with HST for each of the following
           combinations: WFPC-2+F160BW, STIS FUV MAMA+SrF$_2$, and ACS/SBC. \\ 
        6:  Numbers are calculated for the high resolution imager; the wide
  		field imager has a much smaller $\theta$.
        7:  Assumes  2$\times$ stars per magnitude w.r.t. TD-1. \\
	8: Assumes 5000 independent pointings to end-of-life. \\
        9:  m$_L$ transformed from AB; data refers to the all-sky 
		imaging survey (AIS) phase. \\
	10: Number of sources in the 2nd EUVE catalog.
     \end{table}  

    For MSX/UVISI the sensitivity is that of the high resolution imager.
    A full analysis of this data set is not yet available. For the purpose
    of comparison, I assumed that UVISI performed 400 independent
    pointings, yielding a sky coverage of 400 degrees$^2$ with the
    high resolution images and perhaps 4,000  degrees$^2$ with the
    wide field imager.
    The difference in the expected performance of 
    the XMM/OUVM and that of the SRG/TAUVEX lies in the multiple-telescope 
    design of TAUVEX, which allows pure-UV simultaneous observations in 
    three bands, compounded by its larger field of view. The large value of
    $\theta$ for GALEX, despite the similar m$_L$ to that of TAUVEX, results
    from the all-sky coverage. For GALEX, the comparison is done with the
    published expectations for the all-sky survey phase, and the limiting 
    magnitude for 200 nm has been transformed from the AB system to 200 nm
    monochromatic magnitudes.

    Because of the ``unfair'' comparison based on $\theta$, and because not 
    all surveys cover the entire sky, it may be more useful to look at 
    another estimator, the density of sources detected (or which are 
    expected to be detected) by a certain experiment, shown in 
    Figure 5. Expected results from a hypothetical HST survey with the
    ACS/SBC have not been plotted in the figure, but they are expected
    to be significantly deeper than those of GALEX DIS.

\begin{figure}[tbh]
\vspace{10cm}
\includegraphics{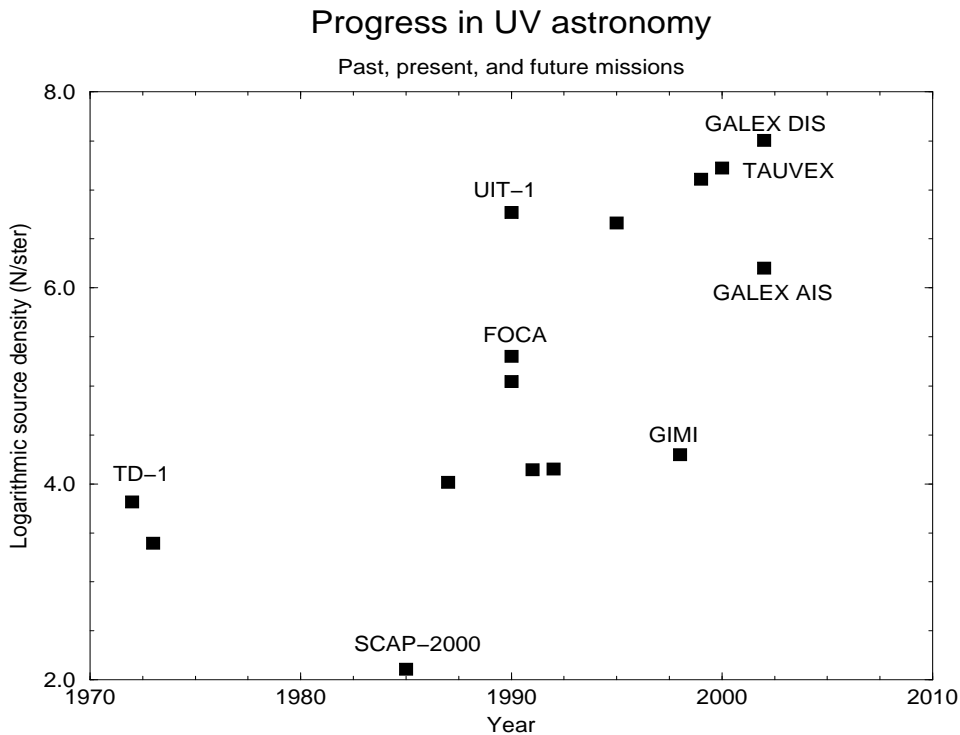}
\caption{Source density in the UV for various sky surveys. All-sky and partial surveys 
(past or future) are included.}
\end{figure}

        \section{The Resultant Sky Picture}

    Any idea we may form of the appearance of the UV sky is necessarily limited
    by the existence of a single all-sky survey in the UV (TD--1)
    and by the deeper surveys in limited regions of the sky. Comparing the number 
    of UV sources charted by TD--1 with similar mapping done in the optical we 
    are forced to
    conclude that the status of UV astronomy is now at par, in the number of
    sources, with the Carte du Ciel
    catalogs from the end of the last century. The importance of knowing the 
    properties of a large number of sources is the astronomical analog of the
    biological ``diversity of species'' argument; the more species are
    identified, the higher is the chance of discovering a new phenomenon. By
    skipping the stage of deep and extensive mapping of the sky in the UV, the 
    astronomers may be missing the discovery of some new types of sources.

    I compared above the status of the UV sky knowledge in the 
    post-TD-1 era to that which prevailed in optical astronomy at the turn 
    of the 20th century. In the same vein, our knowledge of the EUV sky in 
    the post-EUVE era can be compared to a naked-eye look at the sky on a 
    misty night; very few stars, only the brighter or nearer ones, can be 
    seen. The present EUV catalogs are at par, in the number of sources, 
    with the catalogs of Hipparcos (127 BC), Tycho Brahe (1601), and 
    Hevelius (1660). If a full-sky survey could be conducted to the depth of 
    the right angle survey of EUVE, it would presumably record some 50,000 
    sources, about 100 times fainter than those of the "Second EUVE Source 
    Catalog" from the full-sky survey. 
 
    The picture resulting from the all-sky survey of the TD-1 mission, from 
    the EUVE and ROSAT WFC surveys, and from the limited surveys performed 
    previously is that of a UV sky dotted with many stellar and 
    extragalactic sources, where the background is produced by natural 
    emissions (Ly$\alpha$, other emission lines), and/or by UV light 
    scattered off dust clouds. In sharp contrast, the opacity of the ISM is 
    such that in the EUV range, only the more intense and/or nearby sources 
    are detectable. 
 
    In this section I discuss the accumulated knowledge on different types 
    of celestial objects and on the diffuse background, both for the UV and 
    for the EUV ranges. The various sources have been listed in Table 1.
    The main conclusions of this section are, for the UV, that (a) most
    of the brighter UV sources are early-type stars, where the spectral
    type contributing mostly depends on the UV band of the observation, (b)
    at faint 200 nm magnitudes most sources are galaxies, (c) the UV background 
    is mostly starlight from within the Milky Way, scattered by interstellar
    dust, (d) the small fraction of extragalactic UV background can be
    explained by integrated light from galaxies. For the EUV, the conclusions
    are that (a) most of the sources are late-type, coronally-active stars, and
    (b) there are only very few extragalactic sources, presumably because
    of the opacity of the ISM and the intrinsic source opacity for EUV photons.
 
        \subsection{What is known about stars ? } 
 
    The various experiments construct a picture in which most of the stars 
    detected by TD-1, FAUST, SCAP and FOCA are relatively early-type B, A 
    and F. However, most of the stars included in the UIT catalog are 
    probably late-type (G and later). Specifically for TD-1 (Thompson {\it 
    et al.} 1978) the spectral classes B, A, and F make up 95.9\% of all 
    the stars in the published version of the catalog. Carnochan \& Wilson
    (1983) found many TD-1 stars with more intense UV emission than B8 stars
    (m$_{156.5}$-m$_{274}\leq$--1.30). They identified these objects as 
    unreddened subdwarfs, with a scale height in the Milky Way similar to that
    of the central stars of planetary nebulae.

    I show in Tables 4, 
    5, 6, and 7 a comparison of the TD-1 spectral distribution with that from 
    the FAUST observations of the North Galactic Pole, Virgo, and Coma 
    regions. The column labeled HES indicates the number of hot evolved stars
    (horizontal branch, subdwarfs, white dwarfs) in each magnitude bin. Note that
    while the FAUST statistics refer to the sources detected at 165 nm, the
    TD-1 values are for sources fulfilling the selection criteria for 
    catalog inclusion, in particular a S/N$\geq$10 at 156.5 nm.

    \begin{table}[htb]  
 
    \begin{center}  
 
    \caption{UV sources in the TD-1 catalog}  
 
    \begin{tabular}{ c c c c c c c c c}  
    \hline  
    UV mag & O0--O9 & B0--B9 & A0--A9 & F0--F9 & G0-K9 & M0--M9 & Other & 
    Total\\  
    \hline   
    $\leq6.9$  & 34 & 2132 & 2998 & 1886 & 1033 & 19 & 2 & 8104 \\  
    7.0--7.9   & 22 & 2411 & 4397 & 2388 & 41 & 0 & 10 & 9269 \\  
    8.0--8.9   & 24 & 3185 & 6260 & 611 & 13 & 1 & 37 & 10131 \\  
    $>$9.0  & 17 & 1757 & 1928 & 12 & 5 & 0 & 115 & 3834 \\ \hline  
    Total      & 97 & 9485 & 15583 & 4897 & 995 & 20 & 138 & 31338 \\  
    \hline  
    \end{tabular}  
 
    \end{center}  
 
     \end{table}

     \begin{table}[htb] 
    \begin{center} 
    \caption{UV sources in the North Galactic Pole region (l$\approx$281, b$\approx$84)} 
    \begin{tabular}{ c c c c c c c c} 
    \hline 
    UV mag & B0--B9 & A0--A9 & F0--F9 & G0-K9 & HES & 
    AGN/galaxies & Total \\ \hline  
    $\leq6.9$  & 0 & 3 & 0 & 0 & 1 & 0 & 4\\ 
     7.0--7.9  & 0 & 1 & 0 & 0 & 1 & 0 & 2 \\ 
    8.0--8.9   & 0 & 2 & 1 & 0 & 1 & 0 & 4 \\ 
    9.0--9.9   & 1 & 3 & 3 & 1 & 0 & 0 & 8 \\ 
    10.0--10.9 & 0 & 4 & 5 & 0 & 3 & 2 & 14 \\ 
    11.0--11.9 & 1 & 5 & 5 & 1 & 2 & 3 & 17 \\ 
    12.0--12.9 & 1 & 2 & 7 & 3 & 2 & 2 & 17 \\ 
    13.0--13.9 & 1 & 3 & 5 & 3 & 0 & 1 & 13 \\ \hline 
    Total      & 4 & 23 & 26 & 8 & 10 & 8 & 79 \\ 
    \hline 
    \end{tabular} 
    \end{center} 
     \end{table} 
 
    \begin{table}[htb] 
    \begin{center} 
    \caption{UV sources in the Virgo region (l$\approx$279, b$\approx$77)} 
    \begin{tabular}{ r c c c c c c c c c} 
    \hline 
    m$_{UV}$ & B0--B9 & A0--A9 & F0--F9 & G0--G9 & HES & AGN/Galaxies 
    & No ID & No Sp & Total \\ \hline   
    5.0--5.9 & 0 & 2 & 0 & 0 & 0 & 0 & 0 & 0 & 2 \\ 
    6.0--6.9 & 0 & 5 & 0 & 0 & 1 & 0 & 0 & 0 & 6 \\ 
    7.0--7.9 & 1 & 3 & 0 & 0 & 0 & 0 & 0 & 0 & 4 \\ 
    8.0--8.9 & 0 & 4 & 0 & 1 & 1 & 0 & 0 & 0 & 6 \\  
    9.0--9.9 & 2 & 5 & 2 & 0 & 0 & 0 & 1 & 0 & 10 \\  
    10.0--10.9 & 3 & 2 & 2 & 0 & 2 & 6 & 2 & 0 & 17 \\  
    11.0--11.9 & 3 & 6 & 14 & 0 & 2 & 16 & 2 & 0 & 43 \\  
    12.0--12.9 & 4 & 8 & 8 & 1 & 0 & 31 & 3 & 2 & 57 \\  
    13.0--13.9 & 2 & 5 & 6 & 0 & 1 & 20 & 6 & 0 & 40 \\  
    14.0--14.9 & 0 & 2 & 0 & 0 & 0 & 3 & 0 & 0 & 5 \\  
    $\geq$15.0 & 0 & 1 & 0 & 0 & 0 & 0 & 0 & 0 & 1 \\ \hline  
    Total & 15 & 43 & 32 & 2 & 7 & 76 & 14 & 2 & 191 \\ \hline  
    \end{tabular}  
    \end{center}  
    \end{table} 
 
    \begin{table}[htb] 
    \begin{center} 
    \caption{UV sources in the Coma region (l$\approx$60, b$\approx$88)} 
    \begin{tabular}{ r c c c c c c c c } 
    \hline 
    m$_{UV}$ & B0--B9 & A0--A9 & F0--F9 & G0--G9 & K0--M9 & HES 
    & Total \\ \hline   
    $<$7     & 0 & 1 & 0 & 0 & 0  &  0   & 1 \\
   7.0-7.9 & 0 & 1 & 0 & 0 & 0  &  0  & 1 \\
   8.0-8.9 & 0 & 1 & 0 & 0 & 0 &   1   & 2 \\
   9.0-9.9 & 1 & 1 & 0 & 0 & 0 &   2 & 4 \\
10.0-10.9  & 0 & 4 & 3 & 1 & 0 &   3 & 11 \\
11.0-11.9  & 0 & 5 & 7 & 1 & 0 &       2 & 15 \\
12.0-12.9  & 0 & 5 & 2 & 1 & 0 &   3 & 11 \\ 
13.0-13.9  & 0 & 0 & 0 & 0 & 1 &   0  & 1 \\  \hline
Total   & 1 &   18 &   12 & 3 & 1 &   11 & 46 \\  \hline 
    \end{tabular} 
    \end{center} 
     \end{table} 

    A comparison of the tables  shows that at high $\mid$b$\mid$ there 
    are few B stars, thus the A and F types dominate the source counts in 
    the FAUST data. The Coma region shows an unexpectedly high number of hot
    evolved stars; these are probably related to the high galactic latitude 
    open cluster Mel 111 (Brosch \etal 1998). 
 
    Except for the FAUST fields studied at Tel Aviv University (Brosch \etal 
    1995, 1998), most surveys used exclusively correlations with existing 
    catalogs to identify sources. In the fields of NGP-NB and Virgo, where 
    the reduction and identification processes are complete, we find almost 
    equal fractions of A-F stars (70 and 75\%). The only disturbing fact is 
    that we identify no white dwarfs in the three Virgo fields, whereas 
    about 7 are expected; it is possible that these hide among the small 
    fraction of unidentified sources, or that they were identified as 
    ``normal'' stars. The Coma field includes the open cluster Mel 111; for
    this reason it is not typical of other high-b fields.
 
    In the EUV range it is possible to compare the different findings of 
    the survey instruments described in the relevant section, with the 
    exception of ALEXIS, whose sources have not yet been collated into a 
    table. I show the ROSAT WFC and EUV lists in Table 8. It is clear that 
    the large majority of EUV sources mapped by the different experiments
    are late-type stars; the number of extragalactic objects is extremely
    limited.
 
     \begin{table}[htb] 
    \begin{center} 
    \caption{Nature of identified EUV sources } 
     \begin{tabular}{ccccccc} 
    \hline 
    Source        & WFC & WFC & EUVE & EUVE & EUVE & EUVE \\ 
    nature        & BSC & 2RE & BSL  & 1st  & 2nd  & ROSAT \\ \hline  
    AGN, QSO      & 7   & 18  & 10   & 9    & 6    & 0 \\ 
    XRB, CVs      & 20  & 10  & 14   & 16   & 15   & 5 \\ 
    O-A-B stars   &  8  & 8   & 13   & 32   & 18   & 43 \\ 
    F-G-K-M stars & 181 & 251 & 172  & 184  & 161  & 411 \\ 
    WDs (PNN)     & 119 & 140 & 117  & 109  & 98   & 27 \\ \hline
    Total         & 335 & 421 & 326  & 350  & 298  & 486 \\
    \hline 
    \end{tabular} 
    \end{center} 
     \end{table}

        \subsection{What is known about galaxies ? } 
 
    The information about galaxies is very sparse and we still lack a large 
    sample of a few 1000's galaxies, from which to perform good statistical 
    studies. 
 
    Significant information on selected objects, mainly on stellar 
    populations and the nature of the ISM, was obtained from IUE spectra 
    (O'Connell 1992). Similar observational data, combining UV with optical 
    and near-IR spectrophotometry through matched apertures, was recently 
    used to derive template spectral energy distributions (SEDs) for various 
    types of galaxies (Storchi-Bergmann {\it et al.} 1994; McQuade {\it et 
    al.} 1995). A combination of IUE observations and data from other UV 
    imaging missions was used to extract ``total'' UV information on 
    galaxies (Longo \etal 1991; Rifatto \etal 1995a, 1995b). 
 
    There is hope to derive the star-forming histories of galaxies through a 
    combination of data from the UV to the near-IR, in the manner of the 
    Storchi-Bergmann {\it et al.} (1994) templates. However, we have 
    recently shown (Almoznino \& Brosch 1997a, 1997b) that such 
    decompositions are not unique, at least for a sample of blue compact 
    dwarf galaxies (BCDs). In order to account for recent star-formation 
    bursts, it is necessary to include information about the ionizing 
    continuum; as the ISM in the Milky Way and in the target object prevents 
    direct observation of the ionizing continuum, this can be derived from 
    H$\alpha$ observations under simplifying assumptions. UV data collected 
    by IUE or UIT are usually at 
    $\lambda>$140 nm. This region contains mainly radiation from A-type 
    stars and requires extrapolation of the stellar population to earlier types, 
    to account for Lyman continuum photons. 
 
    To understand large populations of galaxies, in terms of stellar 
    populations and star formation histories, it is impractical to rely only 
    on the detailed modeling of spectral features in the optical region. 
    One should combine 
    information from many spectral bands, covering a spectral region as wide 
    as possible. 
 
    In the absence of very deep UV surveys in more than a single spectral band, 
    such as those expected to result from the UIT exposures, our information 
    about a significant number of galaxies originates from the SCAP-2000 
    (Donas {\it et al.} 1987) and FOCA (Milliard {\it et al.} 1992) 
    measurements. These consist of integrated photometry at 200 nm of a few 
    hundred galaxies. In the 200 nm band and in the brightness range 
    16.5-18.5 galaxies apparently dominate the source counts. The 
    corresponding blue magnitudes of these galaxies are B=18--20 and their 
    typical color index is $[200-V]\approx-$1.5. Comparing this index with 
    the template spectra of Kinney {\it et al.} (1996), the FOCA galaxies 
    fit the SB2 template, {\it i.e.,} a slightly reddened starburst galaxy. 
 
    The galaxy counts from the FOCA flights originate from the analysis of 
    three high latitude fields covering about 4.5 square degrees, which 
    include galaxy or globular clusters (M3, Abell 2111, and SA 57, which 
    contains a few Coma cluster galaxies). Milliard {\it et al.} (1992) 
    indicate negligible contamination of their galaxy counts by either UV 
    stars or cluster galaxies. In order to reproduce the observed galaxy 
    density distribution, Armand and Milliard (1994) find that a simple 
    transformation to the UV of the known visible galaxy counts is not 
    sufficient. They require a larger contribution by star-forming galaxies 
    in the recent past, such as late-type spirals or blue dwarf galaxies. 
 
    In the field of A2111, Milliard {\it et al.} (1996, private 
    communication) find that the UV galaxies are mostly in the foreground or 
    the background of the cluster, ranging as far as a redshift of 0.68. 
    Note that the Lyman break in the rest frame of a distant galaxy will 
    enter the FOCA window only at z$\approx$1.3. Spectra of some of these 
    optically faint galaxies reveal narrow emission lines, justifying a link 
    with star-forming galaxies. A larger fraction of starforming galaxies
    at high z was claimed also from HST data, {\it i.e.,} the HDF or the MDS.
    However, note that Giavalisco {\it et al.} 
    (1996) claim that not all HST faint galaxies are starbursts or peculiar 
    galaxies; being observed at UV wavelengths in their rest frame their 
    morphological appearance is later than would be determined from optical 
    studies. 
 
    Using the ``field'' galaxy luminosity function in the UV, derived by 
    Deharveng {\it et al.} (1994) from the smooth linking of the projected 
    density of UV galaxies from the FAUST and FOCA counts, it appears that: 
    \begin{equation}  
    log N(m)=0.625 \times m_{200}-9.5  
    \end{equation} 
    There is a faster increase in the number density of galaxies at faint UV
    magnitudes than it is for stars.
    
    The UIT results, derived from the analysis of the 48 independent 
    pointings from the ASTRO-1 mission, indicate that the majority of UV 
    sources recorded by this instrument are stars (Smith {\it et al.} 1996). 
    This is at odds with the analysis of Milliard {\it et al.} (1992). One 
    possible explanation could be the longer wavelength response of UIT; 
    most of the analyzed images were taken through the A1 filter. This filter 
    has a significant color term because of its considerable width, thus its 
    effective central wavelength is 230 nm for early-type stars and about 
    280 nm for G stars. In contrast, the FOCA bandpass is much narrower and 
    better defined. The difference in the amount of recorded stars could 
    then be the result of UIT seeing more late-type stars, to which FOCA 
    would be ``blind''. 
 
    To check this possibility, we calculated models of stellar densities 
    patterned after those of Brosch (1991), with the difference that the 
    present models were calculated for the UIT A1 band. Specifically, we 
    derived an optical-to-UV transformation based on IUE spectra convolved 
    with the A1 filter response, shown in Fig. 6, calculated the 
    cumulative stellar densities 
    in the direction of UIT targets, and compared the expected and the 
    measured source counts and properties, in 
    the deepest UIT exposures (all with the A1 filter and with 600 sec 
    exposure or longer). We found that we can reproduce the total counts, 
    the distribution of star counts by magnitude bins, and the [UIT--V] 
    color distribution with a model cutoff at V$\approx$18.  Results for 
    the UIT field centered at (06:22 ;
    --13:03) are shown in Figures 7 and 8, and the error bars on the
    experimental points originate from an assumption of Poisson statistics 
    in the number counts. It follows that 
    UIT could indeed have detected mostly stars, and that the Galaxy model 
    in Brosch (1991), with its subsequent
    modifications and transformation to the UIT A1 band,  reproduces
    faithfully the observed stellar distribution.

\begin{figure}[tbh]
\vspace{10cm}
\includegraphics{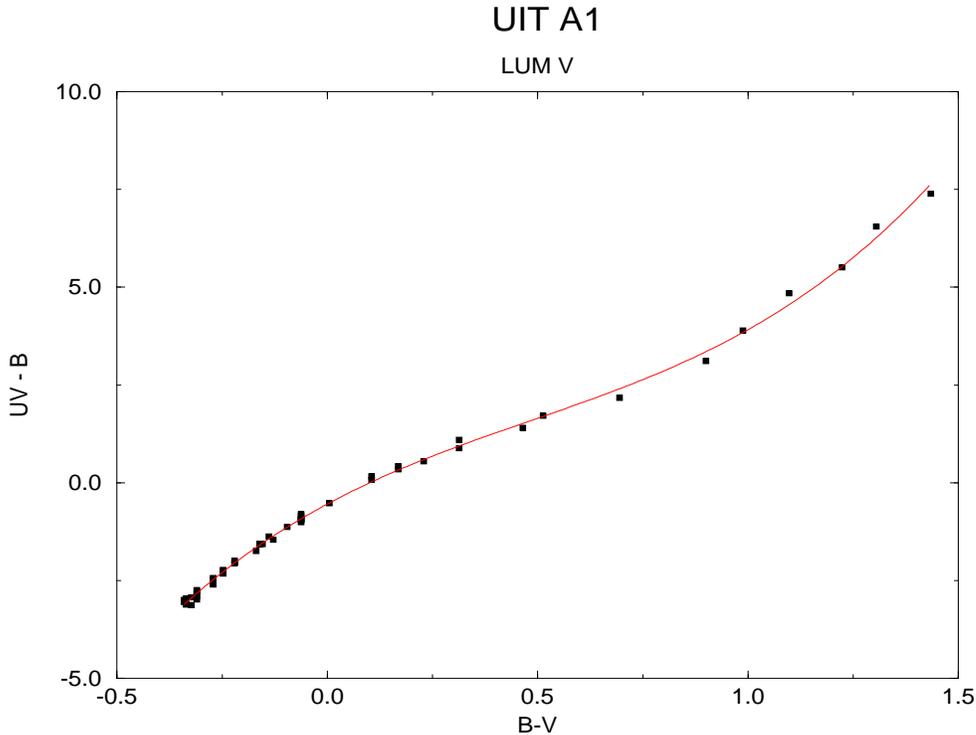}
\caption{Derived transformation between observed B-V of a star and its UV--B
color for main sequence stars. The UV magnitude is derived by convolving 
the UV spectrum (from the IUE observations)
with the transmission of the A1 filter of UIT.}
\end{figure}

    One must be wary of the 
    UV--V colors quoted in the UIT catalog; the V magnitude is derived in many 
    cases from the ``quick-V'' magnitude in the HST Guide Star Catalog, which 
    may be significantly off (0.15 mag=1$\sigma$ for objects near the plate 
    center calibrating sequences, going as high as 0.30 mag=1$\sigma$ for 
    objects far from the sequence: Russell {\it et al.} 1990).

\begin{figure}[tbh]
\vspace{10cm}
\includegraphics{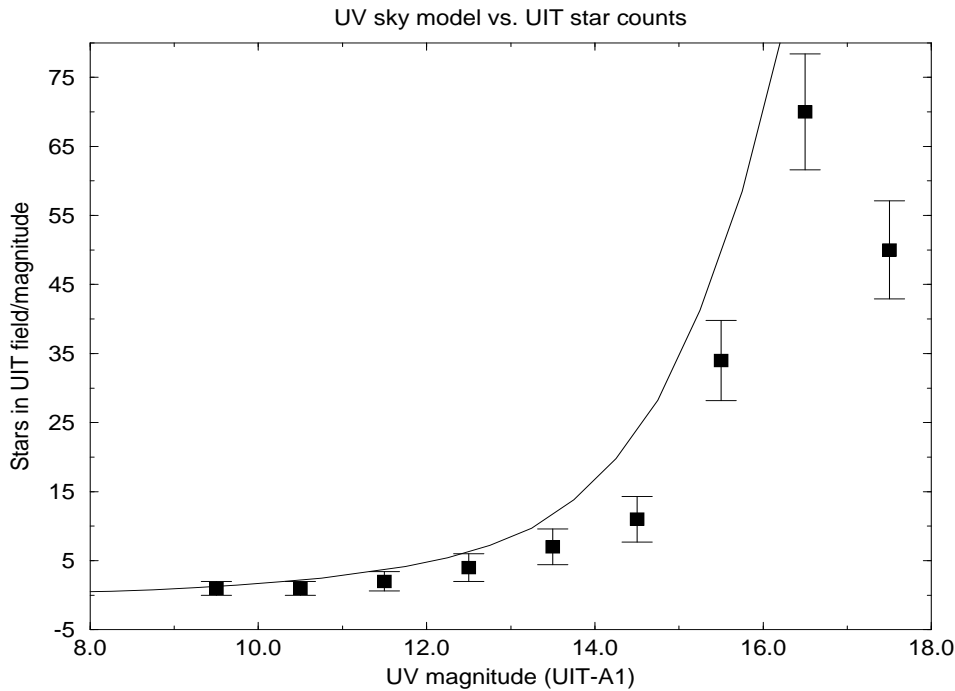}
\caption{Predicted {\it vs.} observed UV star counts at the UIT A1 band, for a
field at (06$^h$22$^m$: -13$^{\circ}$03'). In general, and up to m$_{UV}\approx$16.5,
the model predictions are within one standard deviation of the real star counts. For
fainter stars there is progressive incompleteness.}
\end{figure}

\begin{figure}[tbh]
\vspace{10cm}
\includegraphics{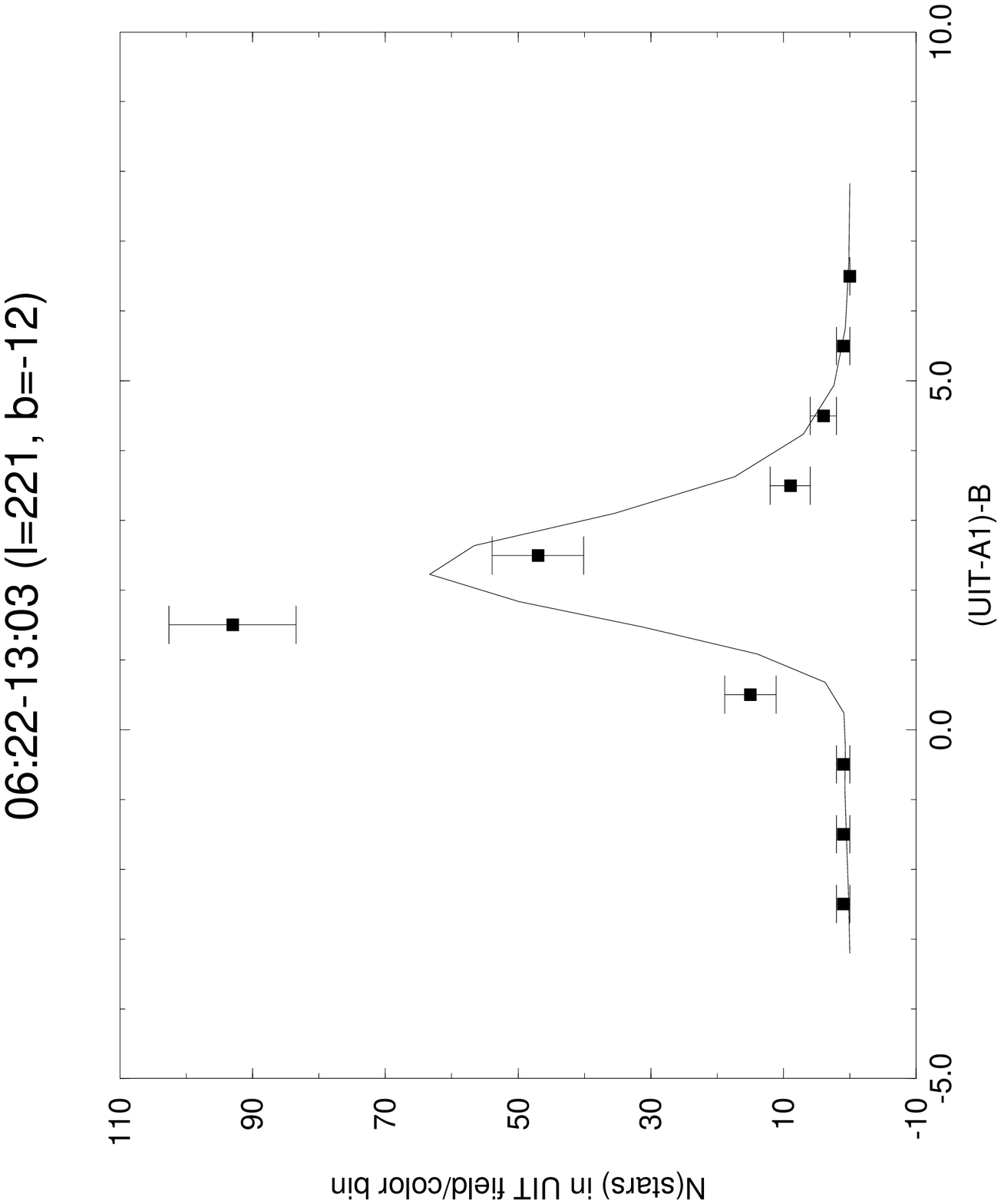}
\caption{Predicted (solid line) {\it vs.} observed (points with error bars)
UV star color distribution, for the UIT field at (06$^h$22$^m$; -13$^{\circ}$03'). 
As in Fig. 7,
most points follow the theoretical color distribution and are  within one standard 
deviation of it. I have no explanation for the outlier point at UV--B=1.5.}
\end{figure}
 
    Another comparison of UIT and FOCA performance is possible using the 
    studies of M51 by both instruments (Bersier {\it et al.} 1994, Petit 
    {\it et al.} 1996 for FOCA; Hill {\it et al.} 1996 for UIT). I measured 
    the integrated flux of 22 HII regions in M51 on the calibrated FOCA image 
    (kindly provided by D. Bersier),
    out of the 28 identified in the UIT B1 image of 
    Hill {\it et al.} (1995). The comparison is shown in Figure 9. In 
    general, the m$_{200}$ and the m$_{152}$ magnitudes correlate well. 
    There is no significant calibration offset between UIT and FOCA, at 
    least in this image, and the HII regions do not show strong reddening 
    effects between 152 nm and 200 nm. This nice correspondence demonstrates 
    fully the advantage of a balloon-borne UV survey telescope, at a 
    fraction of the cost of UIT or of other orbital telescopes (see below). 

\begin{figure}[tbh]
\vspace{10cm}
\includegraphics{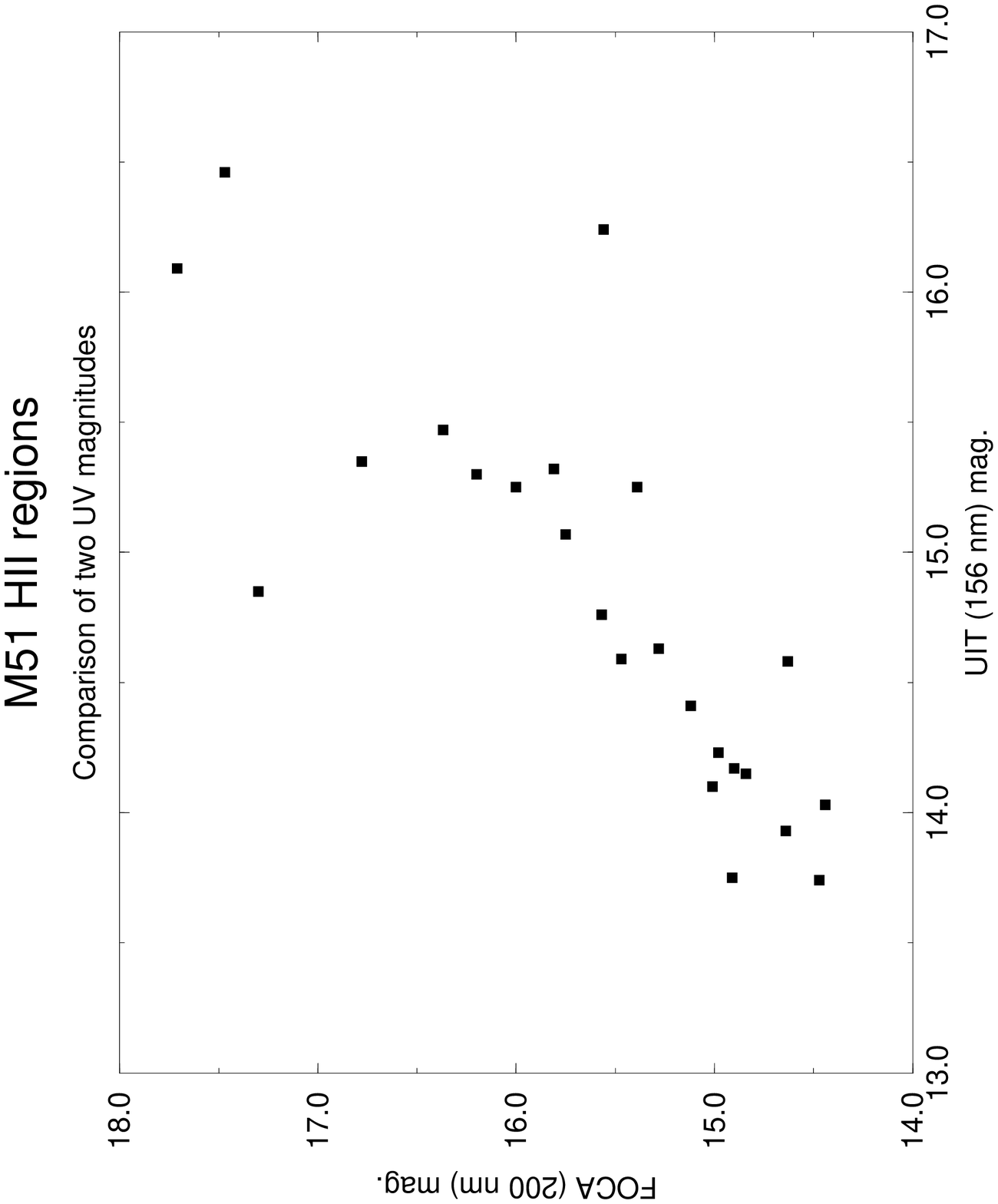}
\caption{Comparison of UIT and FOCA UV photometry of the same regions of M51.}
\end{figure}
 
    One missing ingredient to a fuller model of the sky consists of a proper 
    representation of galaxies and AGNs/QSOs. The latter, in particular, are 
    UV-bright objects. Their detection in deep UV surveys should be 
    relatively easy. Moreover, their spectral energy distribution is very 
    different from that of stars, thus using UV and optical colors it is 
    possible to discriminate them from the foreground stars. The claim by 
    the FOCA group of a large contribution of UV galaxies is supported by 
    theoretical arguments requiring a fast-evolving population of perhaps 
    dwarf galaxies for z=0.2-1.0, in order to explain the faint source counts in 
    other spectral domains (Ellis 1997). A similar possibility, that 
    $\sim$60 of the almost star-like blue objects in the Hubble Deep Field 
    are distant extragalactic sources, has been proposed by M\'{e}ndez \etal 
    (1996). One possibility could be that these sources are distant blue 
    dwarf galaxies or starbursting galaxies, which are smaller than their 
    present-day counterparts. This is apparently confirmed by the Keck 
    spectra of some of these objects, though the inferred star formation
    activity in these objects appears much higher than in local starbursts
    ({\it e.g.,} Guzman \etal 1997). 
 
        \subsection{What is known about the ISM ? } 
 
    Studies by UIT and FAUST emphasize the relative importance of the dust 
    in understanding the UV emission. Bilenko  (1995) analyzed the 
    TD-1 catalog and a version of the Hipparcos Input Catalog transformed to 
    the TD-1 band with information derived from IUE spectra of different 
    types of stars. They showed that the UV exinction is very patchy, with 
    very different values of extinction per kpc on scales smaller than 
    10$^{\circ}$. 
 
    Tovmassian {\it et al.} (1996b) used the results of GLAZAR observations 
    of a 12 degrees$^2$ area in Crux, to establish that the distribution of 
    ISM dust is very patchy, with most of the space between stellar 
    associations being relatively free of dust. They did not attempt to 
    delimit better the location of dust clouds in 3D space, because of the 
    rather sparse sky coverage of the GLAZAR stars. 
 
    Much of our present understanding of the structure of the local ISM 
    comes from studies in the EUV range. In particular, the various EUVE 
    catalogs (Bowyer {\it et al.} 1994, 1996; Lampton {\it et al.} 1996) 
    confirm the previously known features of the local ISM (a ``tunnel'' to 
    CMa with very low HI column density to ~200 pc. and close to the 
    Galactic plane first identified by Gry \etal (1985) from the COPERNICUS
    data, a cavity connected with the Gum Nebula in Vela, a 
    shorter 100 pc. tunnel to 36 Lynx, and the very clear region in the 
    direction of the Lockman hole). It is clear that the present EUV 
    catalogs are not deep enough to reveal finer, or deeper details. 
 
    The new UIT study of M51 shows that $\frac{A_\lambda}{E(B-V)}$ changes 
    with decreasing metallicity (or galactocentric distance). This is 
    reminiscent of the finding by Kiszkurno-Koziej \& Lequeux (1987), that the 
    extinction law in the Milky Way changes with distance away from the galactic 
    plane. The authors  
    find that the H$\alpha$ flux is depleted in the inner regions of M51;
     this they interpret as increased Lyman continuum extinction
    in the inner parts of M51.
 
        \subsection{What is known about the UV and EUV backgrounds ? } 
 
    The accurate measurement of the UV sky background (UVB), with the 
    expectation that it could put meaningful cosmological limits, has been 
    the goal of many rocket, orbital, and deep space experiments. Many 
    observational results were summarized in reviews by Henry (1982), Bowyer 
    (1990), Bowyer (1991),  and Henry (1991).
    It is worth noting the many 
    pitfalls of accurate UV background measurements: sensitivity to 
    instrumental dark current, atmospheric emissions at low orbital and 
    rocket altitudes, {\it etc.}  Leinert \etal (1998) discuss
    observations and interpretation relevant to the UV and EUV backgrounds 
    in a general discussion of the sky background in many wavelengths. 

    Bowyer (1991) separates the various origins of the diffuse UVB into 
    ``galactic'' and ``high latitude''. The latter forms an $\sim$uniform 
    pedestal, onto which the galactic component is added in various amounts 
    depending on the direction of observation. The ``galactic'' component can 
    be $\sim$one order of magnitude more intense than the ``high latitude''
    one. Most of this intensity is probably from light scattered off dust 
    particles in the ISM, and the rest is from the gaseous component of the 
    ISM (HII two photon emission and H$_2$ fluorescense in molecular 
    clouds). The majority of the ``high latitude'' component is probably also 
    galactic, originating from light scattering off dust clouds at high 
    galactic latitude. 
 
    An analysis of the diffuse UV emission (Murthy \& Henry 1995) indicates 
    that the extragalactic component of the UV background can be at most
    100-400 photon units. Whenever the column density of HI 
    is larger than 2 10$^{20}$ cm$^{-2}$, the main contributor to the UVB   
    originates from dust-scattered starlight. The low levels of the extragalactic 
    component are produced probably by the integrated light of galaxies (Armand 
    {\it et al.} 1994), or the integrated light of the Milky Way 
    scattered off dust grains in the Galactic halo (Hurwitz {\it et al.} 
    1991). Even the Lyman $\alpha$ clouds in intergalactic space may 
    contribute, through their recombination radiation (Henry 1991). 
 
    The confirmation that most UVB originates from dust-scattered light was 
    obtained by Sasseen {\it et al.} (1995) and Sasseen \& Deharveng (1996). 
    Sasseen {\it et al.}  showed that the spatial 
    power spectrum of the diffuse UVB on FAUST images is consistent in 
    spectral index and amplitude ratio with the IRAS 100 $\mu$m signal in 
    the same sky areas, while Sasseen \& Deharveng found that the UV
    background correlates with the FIR measurements of COBE/DIRBE. 
    Galactic cirrus clouds apparently scatter back 
    significant FUV radiation from the Milky Way; Harkala \etal (1995) found 
    such UV emission with FAUST from the direction of a dark cloud detected 
    at 100 $\mu$m by IRAS.  
 
    The low levels of UV background, away from orbital and galactic 
    contaminants, have been confirmed by an analysis of 489 UIT 
    images from the ASTRO-1 flight (Waller {\it et al.} 1995). After 
    correcting for orbital background and zodiacal light, and after 
    accounting for scattered Galactic light by ISM cirrus clouds (through
    the 
    IRAS 100 micron emission), the authors extrapolated the UV-to-FIR 
    correlation to negligible FIR emission to find that the extragalactic 
    (cosmic) UV background must be 200$\pm$100 count units. Note one 
    effect in all Shuttle-based experiments, which was not accounted for in 
    the UIT UVB analysis; the influence of background light produced by the 
    Shuttle attitude jets firing. During the FAUST flight this was observed 
    as an enhanced background in the photon stream, which was $\sim4\times$ 
    higher than the level from all other sources (Sasseen 1996, private 
    communication). As this is diffuse light, the influence is clearly 
    dependent on the solid angle of observation and on the size of the 
    aperture. For UIT, this should have resulted in a level $\sim$2.5\% that 
    of FAUST. Because of this, the minimal UVB levels estimated by
    Waller {\it et al.} 
    (1995) must be considered upper limits.  

\begin{figure}[tbh]
\vspace{10cm}
\includegraphics{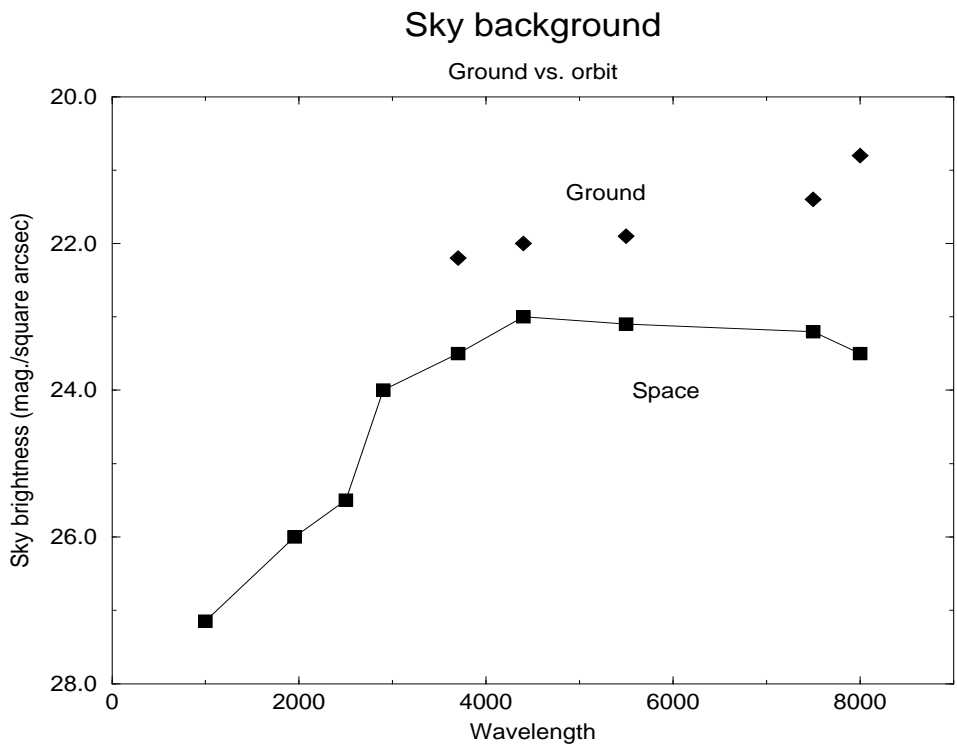}
\caption{UV and optical sky background from orbit {\it vs.} ground observatories.
This is a revised version of a similar figure by O'Oconnell (1987), where the
values related to the optical domain have been retained.}
\end{figure}
    
    The shorter wavelength background has been observed by {\it e.g.,} 
    Holberg (1986) with the Voyager UVS, reaching as far down as 50 nm. The 
    UV range shortward of Ly$\alpha$ but longward of the Lyman break was 
    studied from an integrated data set of all the Voyager UVS observations 
    (Murthy {\it et al.} 1996). They identified 272 UVS observations of 
    which $\sim$50\% have intensities under 100 count units. In particular, 
    note the observations on the North Galactic Pole region, which give 
    consistently a zero background level (upper limit 100 c.u., 1$\sigma$-to 
    be revised). Murthy {\it et al.} concluded, after accounting for the
    scattered light in the Ly$\alpha$ and Ly$\beta$ line wings and for
    the radiation-induced background from 
    the radio-isotope thermal generator on-board the Voyager spacecrafts,
    that the minimal sky background between 
    91.2 nm and 115 nm is 0$\pm$100 c.u. (1$\sigma$). This is, by far, the 
    lowest value of any extragalactic UVB component, and the formal 1$\sigma$
    upper limit is the value plotted in Fig. 10. Its implication is that 
    in the nearby Universe, to z$\approx$0.2, there is only a very small 
    contribution of photons with $\lambda>$912\AA\,. It is also clear that 
    galaxies at any redshift are rather opaque to Lyman continuum photons; 
    these could contribute in the Voyager band from any (reasonable) 
    redshift.
 
    It is possible to evaluate the level of the ``true'' extragalactic 
    background which would be detected by instruments with moderate angular 
    resolution, from the FOCA galaxy counts at 200 nm (Milliard {\it et al.} 
    1992). Their actual galaxy counts, for 15.0$\geq m_{200}\geq$18.5, give 
    (when extrapolated to m$_{200}$=20.0) a contribution of $\sim$100 c.u.'s 
    already
    from the UV-detected galaxies. Note that such faint objects would hardly 
    be detectable in the UIT exposures. Armand \etal (1994) used the FOCA 
    counts at 200 nm, with a limiting magnitude of 18.5, to calculate the 
    contribution to the UV background from galaxies. They estimated that 
    40--130 c.u. can be due to galaxies. A similar analysis by Martin (1997), 
    using FOCA galaxies, indicates that at least 25\% of the UV background can be 
    due to galaxies.  

    Treyer (1997, private communication) calculated the
    expected UV background due to galaxies. This was done by integrating
    the present-day 200 nm luminosity function, obtained from the FOCA 
    observations, to high redshifts. Under the assumption of ``no evolution'',
    she finds that the diffuse 200 nm surface brightness should be 
    $\sim10^{-9}$  erg s$^{-1}$ cm$^{-2}$ \AA\,$^{-1}$ ster$^{-1}$,
    which is equivalent to a surface magnitude $\sigma_{200}\approx$28 mag
    arcsec$^{-2}$, or $\sim$100 c.u. This must be considered a lower
    limit to the background produced by galaxies, as evolutionary considerations,
    such as having more star formation at high redshifts ({\it e.g.,}
    Madau \etal 1996), would increase the value of the EUB.
 
    Any reasonable spectral energy distribution one would invoke for the 
    FOCA UV galaxies should have a SED rising into the UV, thus even higher 
    flux levels are expected in the 100 nm region than at 200 nm where FOCA 
    observes. The Voyager results of Murthy {\it et al.} are in strong 
    conflict with the assumption that the differential count numbers of UV 
    galaxies extend unmodified to m$_{UV}$=24, and in mild conflict 
    (2$\sigma$) with the assumption that they extend to m$_{UV}$=20. In the 
    latter case, {\bf the entire} extragalactic UVB could be due to 
    unresolved galaxies, leaving no room for hydrogen-recombination photons 
    from the Ly$\alpha$ clouds, photons scattered off extragalactic dust 
    clouds, etc. The result of Treyer presented above indicates a difficulty
    with the longer wavelength UVB as well; the contribution by galaxies
    accounts for almost the entire UVB measured by UIT !
 
    A very deep EUVE spectroscopic observation of a large region on the 
    ecliptic has recently been reported (Jelinsky {\it et al.} 1995). It 
    indicates that the only emission lines observed are He I and He II 
    (58.4, 53.7, and 30.4 nm), which originate from scatted Sunlight by the 
    geocoronal and/or interplanetary medium. The spectrum of the EUV 
    background (Fig. 2 in Jelinsky {\it et al.}) shows a tantalizing 
    continuum, which decreases from 800 c.u. at 16 nm to 100 c.u. at 35 nm. 
    This is probably grating scattering of the 58.4 and 30.4 nm helium lines 
    (Jelinsky, private communication). It is expected that the knowledge  
    of the EUV background will improve, as more data from the full ecliptic 
    survey (3 
    10$^6$ sec exposure) will be processed. The lowest value for the EUV 
    background originates from a ROSAT observation (Barber {\it et al.} 
    1996). The detection of the shadow cast by the nearby galaxy NGC 55 on 
    the diffuse 250 eV (49.6 nm) background sets a level of the EUV 
    background coming from extragalactic sources at 29.4$\pm$7.2 keV 
    cm$^{-2}$ sec$^{-1}$ keV$^{-1}$ ($\approx$9.5$\pm$2.3 c.u. at 49.6 nm).

        \section{Future missions }  
 
    The future development of UV Astronomy, at least in the context of a 
    full sky survey, has been influenced strongly by the decisions of two 
    major committees in the US. The first is the ``Decade Survey of Astronomy 
    and Astrophysics for the 1990s'', a committee formed in 1989 by the 
    National Research Council in the US and chaired by John Bahcall to set 
    priorities in an age of diminishing funds. The workings of the Decade 
    Survey committee have been described by Bahcall (1991) and the report 
    itself was published in 1991.  
 
    The Bahcall committee attempted to prioritize by ``democratic consultation'' 
    throughout the astronomical community. It is interesting that among the 
    large and moderate programs recommended by the Bahcall committee there 
    is only one program in the UV domain (FUSE, see below), and the emphasis 
    is on infrared and optical programs. The sociological trend continued 
    well into the small programs, which the committee did not specify beyond 
    a few examples; none of these relate to UV astronomy.  
 
    The second very influencial committee was charted by the Association of 
    Universities for Research in Astronomy (AURA) with support from NASA, 
    and is called ``HST and Beyond''. The committee was formed in 1993, was 
    chaired by Alan Dressler, and produced its report in 1995 (Dressler 
    1996). The Dressler committee identified two goals for the Space 
    Astronomy of the early-21st century: the detailed study of the birth and 
    evolution of normal galaxies, and the detection of Earth-like planets 
    around other stars and the search for life on them.  
    The goals identified by the Dressler committee were to be realized by 
    the Next Generation Space Telescope (NGST), by a space interferometer, 
    and by extending the operation of the HST beyond 2005. In particular, 
    note that HST was to be the major instrument for UV astronomy beyond 
    that date.  
 
    The decisions of the Bahcall and Dressler committees 
    have been adopted by NASA and by the NRC, which control the major 
    funding in the US. In particular, NASA headquarters has been 
    restructured into four major ``themes'', of which the two most related to 
    space astronomy are ``Structure and Evolution of the Universe'', and 
    ``Search for Origins''. It is possible that these decisions have also had 
    far-flung influences beyond the borders of the US, and that other Space 
    Agencies are prioritizing their programs in a hidden race against the US 
    Space Program. This could have had the effect of blocking UV astronomy, 
    at least in the direction of conducting a new all-sky survey, by the 
    major players in space astronomy. 
 
    In the immediate future, only one space mission (GIMI, see below) 
    has the potential 
    of yielding results relevant to the derivation of a full UV sky 
    survey.   GIMI will reach only 
    the relatively brighter objects, an extension of a 
    few magnitudes below the TD-1 limit. With the exception of nearby 
    galaxies, no extragalactic objects are expected to be measured by   
    it. In addition, there is the FUSE mission, which will hopefully begin 
    to operate within 1998. 

    Given these rather pessimistic constraints,  it is 
    gratifying that NASA selected in late-1997 the GALEX mission (see
    below), which was 
    mentioned above in the context of a comparison of surveys. GALEX will
    conduct an all-sky UV survey, comparable to that of the Palomar Sky Survey
    in the optical domain. The results from GALEX will become available in the
    second half of the next decade.

    Before describing the future UV missions, I review a few examples 
    of missions which did not 
    succeed to launch. 
 
        \subsection{Possible missions} 
 
    A number of missions to map the sky to depths comparable with the 
    Palomar Sky Survey (PSS) in the optical have been proposed, but none 
    bar one was 
    accepted by the suitable funding agency. An early attempt was to create a 
    geo-synchronous updated UIT (Largefield Ultraviolet Explorer=LUX), proposed
    in 1986. Another proposed attempt was to have 
    a SMEX for a UV imaging and spectroscopy all-sky survey (PAX, JUNO), then a 
    MIDEX for a similar purpose (MUSIC). These proposals, although very good, 
    did not survive the NASA selection. A third attempt, GALEX (from the
    same PI as PAX, JUNO, and MUSIC), was finally selected in 1997 for a 2001
    launch.

    A mission to study the diffuse UV
    background (HUBE) was retained by NASA as backup for a selected mission.
    A Canadian group proposed to study a 
    mission similar to JUNO for a National Canadian small satellite. Some of its 
    members were among the initiators of STARLAB, a one-meter class Space 
    Schmidt proposed for UV surveys, which was scrapped in the late-80s. It
    is possible that the Canadian option will be to fly on a Shuttle-carried long
    duration platform.
 
    A conceptual design for a very ambitious survey telescope, dedicated to 
    UV astronomy, was presented by the Byurakan Observatory (Armenia) and by the 
    Astronomical Institute of Potsdam, Germany (Tovmassian \etal 1991d). The 
    Astrophysical Schmidt 
    Orbital Telescope (ASCHOT) would be able to image a 2$^{\circ}$.5 field 
    of view with 2" resolution, using an all-reflecting Schmidt design with 
    an 80 cm aperture. The calculated performance indicated a detection 
    limit with S/N=10 at m$_{150}$=24.0 in a 30 min exposure. However, there are 
    formidable aspects in dealing with a 9000$\times$9000 pixel array.
    The primary mirror (diameter 120 cm) and the reflective corrector plate 
    (diameter 80 cm) of ASCHOT have been fabricated by Carl 
    Zeiss Jena prior to the re-unification of Germany. The realization of
    the ASCHOT project 
    remains a possibility worth exploring.  
 
    In Russia Boyarchuck and collaborators are building Spectrum UV (SUV), 
    an HST-like mission intended for deep spectroscopy in the UV of selected 
    targets. Note that this mission, third in line among the SPECTRUM 
    spacecraft, will not be a survey instrument but will be dedicated to 
    studies of known celestial objects. The SUV mission will be almost 
    exclusively dedicated to intermediate resolution spectroscopy, but its
    focal plane camera has imaging possibilities similar to those of
    the  HST ACS (B. Shustov, 1997 private communication). Also, given
    the delays in launching the first Spectrum mission (Spectrum X-$\gamma$),
    it is doubtful whether SUV will be realized before the year 2000.
 
       \subsection{Real missions }  
 
    The one space experiment constructed exclusively for UV astronomy is FUSE 
    (the Far Ultraviolet Spectroscopic Explorer mission), manifested for
     a late-1998 launch. This satellite is 
    desiged to provide high spectral resolution observations 
    ($\frac{\lambda}{\Delta\lambda}$=30,000) in the band 90.5 to 119.5 nm. 
    FUSE is designed for a three year life in low Earth orbit and has 
    significant effective area peaking at $\sim$100 cm$^2$ near 105 nm. This 
    allows high resolution spectroscopy with S/N=20 of objects with 
    monochromatic magnitude 12 in 100,000 seconds of observation, which is 
    equivalent to about 3.3 days, given the expected operating efficiency. 
    Most of the observing time will be dedicated to a number of key program, 
    such as the study of deuterium abundance in the interstellar and 
    intergalactic space.  
 
    The GIMI imaging instrument is integrated into 
    its carrier spacecraft (ARGUS, from the USAF). It is primarily 
    military in character aiming at detecting and 
    characterizing sources of UV emission (or atmospheric opacity), which 
    could affect the detection, identification, and tracking of missiles and 
    warheads. ARGUS is expected to be 
    launched in late-1998 on a Delta II vehicle. 
 
    The GIMI instrument on ARGOS has as one of its 
    declared goals the production of a full sky survey in three UV bands. The 
    most recent description of GIMI is by Carruthers \& Seeley (1996). GIMI 
    consists of two bore-sighted EBCCD cameras, each imaging a 
    $10^{\circ}.5\times10^{\circ}.5$ field of view with a resolution of 
    $\sim$3'.9. The paper gives also the sensitivities expected (Fig. 9). 
    The GIMI imagers cover three spectral regions with two detector-telescope
    assemblies. Camera 1 has a KCl cathode sensitive to the 75-110 nm band 
    and with a nominal field-of-view (FOV) of
    10$^{\circ}.5\times10^{\circ}$.5. Camera 2 has a split FOV: an area of
    7$^{\circ}\times10^{\circ}$.5 is covered by a KBr cathode sensitive to
    the 131-160 nm band, while an adjacent area of 3$^{\circ}.5\times10^{\circ}$.5
    is imaged on a CsI cathode sensitive to the 131-200 nm region.
    The most sensitive arrangement is Cam. 2 with the CsI cathode, whereas 
    Cam. 1, with its KCl cathode, has a sensitivity about 1/20 lower. In its 
    sky survey mode, GIMI will stare for at least 100 sec at each source; 
    from this, its magnitude limit is expected to be $\sim$13.6 in the CsI 
    band. 
 
    The TAUVEX (Tel Aviv University UV Explorer) payload (Brosch {\it et 
    al.} 1994) represents the most advanced attempt to design, build and 
    operate a flexible instrument for observations in the entire UV band.
    The experiment is built for Tel Aviv University by El-Op, Electro-Optical
    Industries, Ltd., with the largest part of the funding from the 
    Government of Israel through the Ministry
    of Science and Arts and the Israel Space Agency. 
    TAUVEX images the same 0$^{\circ}$.9   FOV with three co-aligned telescopes 
    and with an image quality of about 10". It was originally conceived for 
    a small satellite of the OFEQ or SMEX class, but is now part of the 
    scientific complement of the SRG spacecraft scheduled to launch in
    late-1999 or in 2000. 

    The detectors used in TAUVEX are tailor-made by DEP-Delft Instruments
    of Roden, the Netherlands. They consist of CsTe cathodes deposited on 
    CaF$_2$ windows. The photo-electrons are amplified by a three-stack chevron  
    arrangement of MCPs and the detection is by a wedge-and-strip anode. 
    The light passes through filters defining six spectral bands, from a
    very wide one which is essentially a ``blue cutoff'' at $\sim$200 nm,
    to three intermediate ($\sim$300 nm wide) bands spanning the working
    range 135-290 nm, to two ``narrow'' bands at selected regions of interest.
    A view of TAUVEX
    as it would look in space is shown in Fig. 11, where the TAUVEX thermal 
    model (now at Lavotchkin Industries in Moscow) is installed in a solar
    simulator chamber in preparation of a thermal vaccum test. The apertures of
    the three telescopes are visible on the right part of the figure 
    and the telescopes themselves are wrapped in their thermal blankets.

\begin{figure}[tbh]
\vspace{10cm}
\includegraphics{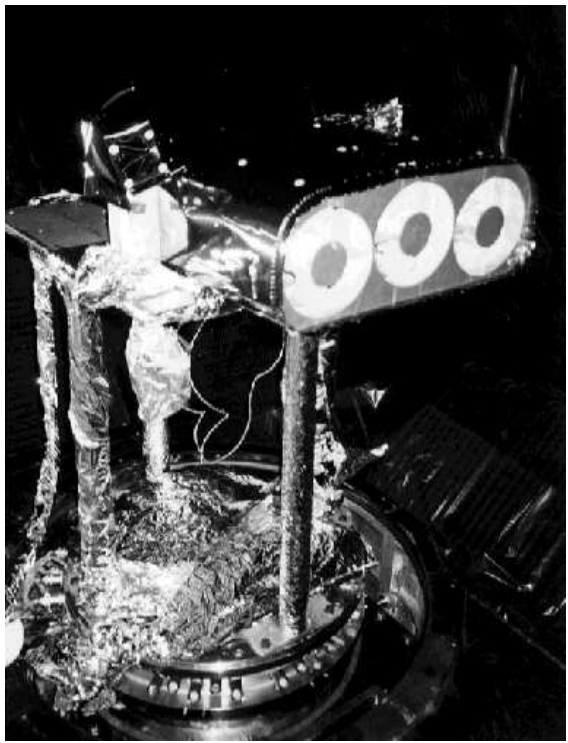}
\caption{The thermal demonstrator model of TAUVEX is identical in external appearance to
the flight model. It is shown here during installation in a space environmental chamber,
prior to qualification under simulated Solar radiation. The three telescope apertures are
visible in the front of the instrument.}
\end{figure}
 
    TAUVEX offers significant redundancy at component and system levels in 
    comparison with other missions. The projected performance is detection 
    of objects 19 mag (monochromatic) and brighter, with S/N$>$10 and in 
    three $\sim$40 nm wide bands, after a four hour pointing. At high 
    galactic latitudes each such pointing is expected to result in the 
    detection of some tens of QSOs and AGNs (mainly low-z objects) and some 
    hundreds of galaxies and stars. It is expected that a three-year 
    operation of TAUVEX on SRG will yield a deep survey of the UV sky to 
    m(UV)=19 mag in at least three UV bands, which will cover 
    $\sim$5\% of the (high galactic latitude) 
    sky. This is based on the size of the field of view and the expectation 
    that the total number of independent pointings after three years in 
    orbit will add up to $\sim$3,000. 
 
    On the same SRG platform, co-aligned with TAUVEX and the rest of the 
    higher energy instruments, will operate also the F-UVITA instrument. It 
    consists of a pair of 20 cm telescopes imaging a 1$^{\circ}$.2 field 
    with 10" resolution in  spectral bands between 91 and 99 nm. Each band 
    is $\sim$9 nm wide. The most up-to-date description of F-UVITA can be 
    found in its homepage at the Paul-Scherrer-Institute in Switzerland,
    where its sensitivity is advertised as adequate to detect V=15 B0 stars.  
 
    Some UV capability has been constructed into the Optical/UV Monitor 
    (OUVM) of XMM (Mason \etal 1996). The OUVM is a 30 cm 
    modified Ritchey-Chr\'{e}tien telescope which 
    images a 17' FOV with $\sim$1" resolution. The light is analyzed
    with filters and with a grism. The UV-optical detector operates
    in the 160-600 nm spectral range and consists of a fast readout 
    256$\times$256 pixel CCD, coupled by fiber boule to the phosphor output 
    of an image intensifier. The CCD readout is analyzed by a 
    transputer-based processor and each photon event is centered to 
    $\sim$0.2 pixel (0.5" for normal operations). The difference 
    from TAUVEX lies in the XMM OUVM being a 
    single telescope time-sharing between the optical and UV ranges, and in 
    the higher spatial resolution, realized at the expense of the smaller FOV. 

    In fall of 1997 NASA announced the selection of a Small Explorer Experiment
    (SMEX) dedicated to a new all-sky survey in the UV. This is the Galaxy
    Evolution Explorer (GALEX: C. Martin P.I.; Bianchi \& Martin 1997) whose 
    prime goal is the survey
    of galaxy evolution effects for the range 0$\leq$z$\leq$2. 
    GALEX is a collaboration of CALTECH, University of California-Berkeley, the
     Johns Hopkins University, Columbia University,
    and the Laboratoire d'Astrophysique Spatiale du CNRS of Marseille, France.

    GALEX is a single
    wide-field imaging telescope with a 50 cm diameter primary mirror and
    a 26 cm diameter secondary. Its focal plane is 1$^{\circ}$.2 wide and
    is equipped with two photon-counting
    position-sensitive detectors. The angular resolution is 3"-5" at
    80\% encircled energy. The light is shared between the two
    detectors by a dichroic beam splitter and a folding flat mirror. The detectors
    cover the spectral region 135-300 nm, selected by detector window
    transmission, cathode response, and
    beamsplitter coatings. The nominal performance of the detectors, which are cathodes 
    with stacked MCPs and crossed, helical delay lines as anodes, allow for
    4096 pixels across the field-of-view. The two GALEX bands, as presently defined
    are 135-180 nm (CsI and CaF$_2$), and 180-300 nm (CsTe and fused silica).

    The GALEX mission is very ambitious, as it has both imaging as well as
    spectroscopic aspects. It shall be conducted from a low Earth orbit, into 
    which the Pegasus launcher will insert the satellite. GALEX will observe by
    scanning the sky during the orbital night. The all-sky imaging survey (AIS) 
    phase is  baselined to last four months,
    after which a two-color catalog of sources to m$_L\approx$19.4 will be
    produced. In addition, it is expected that the sky background will be detected
    by binning the data to 1 arcmin$^2$ and after subtracting the point sources.
    GALEX will then conduct a deep imaging survey (DIS) of 
    $\sim200$ degrees$^2$
    of the sky, which will include areas surveyed by the HDF North and South,
    the ESO Imaging Survey regions, areas of the Sloan survey, {\it etc.} The
    deep imaging survey will detect sources $\sim$4 magnitudes fainter than the
    all-sky survey. Some indication of source variability will be obtained
    from the 80-100 field revisits done during the AIS and DIS phases.

    The spectroscopic survey, to $\sim$1 nm resolution, shall be completed during 
    additional four-month
    periods by inserting a grism in the converging beam, before the beamsplitter. A
    shallow spectroscopic survey to R$\simeq$150 shall be conducted in the 
    first period over 
    the entire region of the DIS, after which two additional periods shall be 
    dedicated to deeper surveys of 20 degrees$^2$, then 2 degrees$^2$, for the
    Medium Spectroscopic Survey and the Deep Spectroscopic Survey, respectively.
    The planned mission allows for a four-month contingency/Associate Investigator
    programs.

    It is now possible to compare the performance of the future missions which
    will bring significant information about the deep UV universe. The three
    experiments, XMM UV/Optical monitor, TAUVEX, and GALEX, reach approximately
    similar limiting magnitudes. GALEX has a significant advantage in its
    global sky coverage, but it is restricted to two simultaneous bands
    {\it vs.} three for TAUVEX. In comparison with the other two instruments, 
    GALEX also does not include a long-term continuous photometric monitoring 
    option. Its spatial resolution is advertised to be similar to that of
    the OUVM, twice better than that of TAUVEX. Both the 
    XMM UV/Optical monitor and TAUVEX shall operate alongside high energy
    imagers, allowing a good determination of the nature of emitting sources.
    Their deployment in high-altitude orbits {\it vs.} a LEO environment for
    GALEX will presumably ensure a better S/N, given a lower sky and
    particle background.
 
        \subsection{Wish list } 
 
    It is clear that ideally ``clean'' observations of the UV sky background 
    should be made far from the Earth's geocorona, only out of the Galactic 
    plane, to avoid the dust-scattered starlight. For practical purposes, we 
    shall have to settle in the foreseeable future with observations which 
    avoid the dust in the Solar System. For this, a telescope must be 
    located beyond the orbit of Saturn, and this requires a specialized 
    instrument. One possibility is to design multi-purpose missions to the 
    outer planets. These, during their cruise phase or when not collecting 
    encounter data, could be used to measure astronomical targets, such as 
    the UV sky background. Toller (1983) and Toller \etal (1987), for 
    instance, used the Pioneer imaging
    photopolarimeter to map the optical emission from the Milky Way in blue
    (395-485 nm) and in red (590-690 nm) from beyond the asteroid belt,
    where the contribution of the zodiacal light was found to be negligible. 

    The findings by Toller were confirmed by the analysis of Schuerman
    \etal (1997), where the boundary beyond which the zodiacal light
    contribution becomes very small, in comparison with the diffuse
    Galactic background, was set at 2.8 a.u. Gordon (1997) performed an
    analysis of the Pioneer 10 and 11 data from beyond this limit and found an
    excess red component over that predicted from the blue light distribution
    (which could be fully attributed to scattering by dust). A location
    beyond the zodiacal dust cloud, therefore, would be ideal for a 
    measurement of the diffuse UVB and of the exceedingly faint UV 
    galaxies. Note also 
    that some missions adopt an ``above the ecliptic'' trajectory to their 
    target; this also would minimize the contamination by zodiacal light
    and offer the deepest astronomical observations. 
 
    Two possibilities, one real and another potential, are to include UV 
    observations during the cruise phases of either the Fast Pluto Flyby 
    (``Pluto Express'') or a possible Neptune orbiter. The Pluto Express 
    mission was being designed when its funding was frozen in 1996 with a 
    probable launch date not before the turn of the century. The Neptune 
    Orbiter has not yet even been proposed officially, but would
    be a logical 
    extension of the Cassini mission. Assuming that Cassini will reach 
    Saturn by 2004, a Neptune Orbiter could launch by 2010 for a Neptune arrival 
    5-10 years after that. Most of its cruise phase could be made available, 
    if so planned, for astronomical observations. Similar possibilities exist 
    for a putative Interstellar Precursor mission.
 
    Another, much cheaper option, altough restricted in its spectral band of 
    observation, is to follow up and expand on the experience gained by the 
    Geneva/LAS group in UV observations from balloons. Observations from 
    stratospheric altitudes are limited to a spectral band $\sim$15 nm wide 
    centered near 200 nm. For a while, it looked like the SR-71's turned to 
    NASA by the USAF could be used for this type of obsevation. However, 
    these planes are again no longer available for scientific research. 
    Despite this limitation, much can be gained from such a mission, if it 
    would rely on the heritage of long-duration balloon flights with updated 
    hardware. 
 
    Super-pressure balloons, flying at 40-45 km altitude, can perform missions of 
    weeks to months. If launched in the Antarctic, the wind patterns keep 
    the balloon mostly over the Antarctic continent. A launch in the Arctic 
    will probably have the same characteristics, but may intrude too much in 
    patterns of civilian and military flights over the North Pole. 
    Trans-oceanic flights at intermediate or equatorial latitudes are also 
    possible and may be more convenient power-wise. Present day technology 
    allows real-time operation of a long-duration balloon mission by using 
    the TDRSS data relay satellites while having a high data rate (Israel 
    1993). Adding solar panels to a regular long-duration balloon is not
    a difficult task, and is one of the possibilities offered in the latest
    (1997) NASA draft call for ``University Explorer'' missions (note though 
    that super-pressure, extremely high altitude balloons are not included
    in the UNEX AO).    Such  a mission is 
    therefore doable and will have a reasonable cost. I thus propose to study the 
    possibility of designing a UV sky survey at 200 nm from a long-duration 
    balloon platform. 
 
    A major cost of any sky survey mission lies in building and qualifying 
    space hardware. These components could have been saved in the cost of such a 
    mission if the GLAZAR-2 telescope could have been reactivated. This requires 
    first an assessment of the status of the optics, the tracking mechanism, 
    and the gimbals which compensate motions of the MIR, which were left 
    attached to the space station for almost one decade with no use. A 
    new focal plane needs to be designed, built, tested, assembled in place 
    of the damaged one, and the new instrument can be activated in space. 
    However, the limitation now lies with the MIR itself, which has 
    surpassed its design lifetime. 
 
    Finally, an interesting possibility is to conduct a deep UV survey of a 
    very small fraction of the sky with the HST. The WFPC-2 has the 
    capability to image $\sim$21 mag UV sources to S/N$\approx$10 with   
    deep observations using the F160BW and/or the F218W filters while
    covering a significant sky region. 
    It is thus possible to design a survey of UV sources as a ``parallel''
    program. A set of such observations could provide information about the 
    faintest UV sources, beyond the capability of UIT, FOCA, TAUVEX, 
    or GALEX, with 
    good morphological information, which could be used to settle the issue 
    of the cosmological UVB. Interestingly, observations with the F160BW 
    filter will be almost devoid of stars ! Our UV-galaxy model, using the 
    bandpass of F160BW, predicts less than one star/square degree (down to 
    V=20) which could be detected in $\sim$1000 sec exposures. 
 
    Using the WFPC-2 exposure time calculator, one finds that a 2,000 sec 
    exposure at high galactic latitude reaches a S/N of 1.8 with F160BW and 
    7.8 with F218W for an unresolved V=20 Sbc galaxy. Adopting the FOCA 
    galaxy count from Armand \& Milliard (1994) and assuming the objects 
    have $\sigma_V\approx$20 mag arcsec$^{-2}$, a 10,000 sec exposure with F218W 
    will yield a S/N$\approx$1 per WFPC-2 pixel. At least one such object 
    should appear on almost any deep F218W exposure. This will be, in
    principle, even deeper 
    with the STIS, once parallel observations with this instrument will be 
    allowed. The HST has therefore the potential of making important 
    observations in the mode of an unbiased survey for faint stars and 
    galaxies. 

A possible instrument studied for the 2002 HST servicing mission is the
Wide Field Camera 3 (WFC-3), which will replace WFPC-2. Its CCDs will
probably have high quantum efficiency in the UV, perhaps an order of
magnitude better than WFPC2. The existence of this camera, with a field
of view of 160"$\times$160" and a pixel size of 0.04 arcsec, will make a
UV survey with the HST even more attractive.

    In the domain of wishes valid for every branch of astronomy one must 
    include a wavelength-sensitive, photon-counting imaging detector. One 
    line of such detectors was described by Perryman \etal (1992, 1993, 
    1994). They consist of superconducting tunnel junctions (STJs) in which a 
    photon creates a cloud of charge carriers, which provide essentially 
    a low resolution imaging spectrometer with
    $\frac{\lambda}{\Delta\lambda}\approx$30. Presumably such detectors 
    would be even more useful in the UV, because of the higher energy of the 
    analyzed photons, allowing a better energy resolution. A single-pixel 
    device of this type was already tested successfully in the laboratory 
    (Peacock \etal 1996) and a quantum efficiency of $\sim$50\% was 
    demonstrated for $\lambda\lambda$ 200-1000 nm, with a spectral 
    resolution of 45 nm. A large-format array of such devices, possibly
    mounted at the WFPC position on HST, would provide an imaging 
    spectrometer with unique properties. This apparently has been
    proposed to ESA as a follow-up instrument for the 2002 HST refurbishing
    mission (Peacock \etal 1997) but was not selected, presumably 
    because the difficulties of a cryogenic 
    system to maintain the detector at T$\leq$1K in the HST environment seen
    daunting.
 
       \section{Conclusion} 

    I have shown here that while UV astronomy flourished in the early and 
    mid-70's, the field stagnated, at least where sky surveys are concerned,
    from then on. In the early days of UV studies, a valuable resource 
    appeared with the production of the TD--1 catalog. This is still,
     20 years after its publication, the only all-sky survey in the
    UV. The TD--1 survey detected mostly stars, but more sensitive
    instruments revealed very interesting extragalactic sources 
    (AGNs, star-bursting
    galaxies, etc.). The most sensitive instrument now available for UV
    imaging is the HST, which can hardly qualify as a survey instrument.
    In terms of surveys, a success of the GIMI and UVISI missions will 
    extend the TD--1 survey by about five magnitudes, and even then,
    only few extragalactic objects would be accessible. The big step
    forward will come with the GALEX, TAUVEX and XMM  UV/Optical monitor
    surveys.

    The EUV range was explored by the ROSAT WFC and by the EUVE, but due
    to the opacity of the ISM, only relatively nearby objects were detected.
    The outstanding sources are viewed through ``windows'' in the ISM, where
    low HI column densities are encountered. At present, there are no prospects
    for an EUVE follow-up mission, but the extension of the EUVE operations
    promises more and interesting data.
 
    It is clear that, with the existing attitude toward pure research and 
    space activities in general, it will be extremely difficult to channel 
    funds for new space missions which will perform additional UV sky surveys
    deeper than and beyond that by GALEX.  
    Therefore, the improvement in human knowledge of the 
    astronomical UV sky can only come about through judicious use of 
    existing, or multi-purpose platforms, where the UV science piggy-backs 
    on other spectral ranges. Not only astronomy missions must be considered 
    as able to yield good UV data, but also planetary or upper-atmosphere 
    science platforms are eligible. Fully new, revolutionary designs tend to be 
    costly; at the present stage they should be avoided. The heritage of 
    past missions should be fully realized before embarking on new 
    adventures. In this light, the extension of the EUVE mission in a 
    low-cost mode is commendable; the continuation of the right-angle surveys 
    should reveal more faint EUV sources. 
 
    Science needs the multi-spectral all-sky UV imaging survey 
    to 19-20 monochromatic magnitude which will be provided by GALEX, 
    as well as a combination of cheaper 
    alternatives, including a long-duration, very high-altitude balloon with 
    a FOCA or ASCHOT-type telescope and a high-resolution detector with 
    electronic readout for a sky survey in the 200 nm band. A mini-survey with 
    HST in the UV, perhaps with an STJ camera, will take advantage of the
    exquisite imaging capabilities of this facilty. An extended EUV 
    all-sky survey, $\sim100\times$ more sensitive than EUVE and with better 
    angular resolution, is also required. The inclusion of a UV sky survey
    phase in planetary missions bound for the outer Solar system, to observe
    from beyond the zodiacal dust cloud, is advocated.
 
        \section*{Acknowledgements} 
 
    UV research at Tel Aviv University is supported by special grants from 
    the Government of Israel the Ministry of Science and Arts, through the 
    Israel Space Agency, and from the Austrian Friends of Tel Aviv 
    University, as well as by a Center of Excellence Award from the Israel 
    Science Foundation. I acknowledge support from a US-Israel Binational 
    Award to study UV sources imaged by the FAUST experiment, and the 
    hospitality of NORDITA, the Danish Space Research Institute, and
    the Space Telescope Science Institute, where 
    parts of this review were prepared. I am grateful for the help of many 
    individuals in producing this review. Benny Bilenko calculated 
    sky models, Liliana Formiggini recalculated the optical-to-UV transformation
    and compared the predicted and actual UV brightness for Virgo,
    Hrant Tovmassian explained intricacies of the GLAZAR series, 
    and Jeff Bloch from the ALEXIS team supplied sky charts and information. 
    Bruno Milliard 
    from LAS Marseille clarified a number of points related to FOCA 
    observations and allowed me to look at a FOCA UV image and optical 
    spectra of a few 
    galaxies in it, Michael Lampton from the Berkeley's SSL/CEA explained 
    intricacies of the EUVE source count, Jesse Hill from the UIT team 
    explained UIT results, David Bersier from the Observatoire de 
    G\'{e}n\`{e}ve 
    added information on M51 and supplied its calibrated FOCA UV image, 
    Jayant Murthy added information about UVISI, Luciana Bianch provided
    GALEX details in advance of publication, and 
    David Israel from GSFC explained the TDRSS-balloon connection; for these 
    I am very grateful. Mark Hurwitz supplied information and a list of 
    publications from the ORFEUS flights; Bill Waller did the same for UIT, 
    read an early version of the paper, and provided some useful remarks. I 
    thank Alan Dressler for some information on his committee's report,
    Marie Treyer for calculating the contribution to the UV background from
    FOCA galaxies, and Prab Gondhalekar for constructive remarks upon reading 
    one of the final drafts. Boris Shustov kindly supplied details about the
    Spectrum-UV mission. I am grateful to the referee for very constructive
    remarks.

    \end{document}